\documentclass{article}
 \pdfoutput=1
\usepackage{graphicx}  
\usepackage{amsmath}   
\usepackage{amssymb}   
\usepackage[export]{adjustbox}
\usepackage{bm} 
\usepackage{dcolumn}
\usepackage{color}
\usepackage{mathrsfs}
\usepackage{amsfonts}
\usepackage{varioref}
\usepackage{physics}
\usepackage[vcentermath]{youngtab}
\usepackage{wrapfig}
\RequirePackage[colorlinks,citecolor=blue,urlcolor=magenta,linkcolor=black]{hyperref}
\usepackage{tikz}
\usepackage{amsmath}
\usetikzlibrary{decorations.pathmorphing}
\usepackage[active]{srcltx}
\usepackage[numbers, sort&compress]{natbib}
\usepackage{tikz} 
\usepackage{subcaption}

\usepackage{float}
\usepackage{placeins}

\title{Luminosity Signatures of Dark Sector Particles from
Black Hole Evaporation in Neutron Stars}

\author{
Ioannis Dalianis\footnote{ntalianis.ioannis@ucy.ac.cy}
 \, and \,
Anastasios Irakleous\footnote{irakleous.anastasios@ucy.ac.cy}
\\[2mm]
\small Department of Physics, University of Cyprus,
Nicosia 1678, Cyprus
}

\date{}

\begin{document}
\maketitle

\begin{abstract}
Microscopic black holes may form at the centers of neutron stars through
the accumulation and collapse of dark matter. While Standard Model
particles produced by Hawking evaporation are efficiently absorbed by the
dense stellar medium, sufficiently weakly interacting and long-lived
particles can escape and subsequently decay into high-energy neutrinos,
photons, or charged particles, providing an indirect probe of Hawking
radiation. We examine the joint energy--angular distribution of these
secondary particles, incorporating mediator propagation, energy-dependent
decay, and relativistic decay kinematics. For relativistic mediators, the
normalized energy-integrated angular profile becomes approximately
independent of the black hole temperature, with its characteristic extent
controlled primarily by the combination $c\tau_S/D$ of the mediator
lifetime and source distance, as the boost enhancement of the decay length
is compensated by relativistic beaming.
We demonstrate the viability of this mechanism and illustrate
its phenomenology with gravitationally coupled scalars,
axion-like particles (ALPs), dark photons, and heavy neutral leptons.
Comparing with monochromatic mediator production from dark matter
annihilation, we find that the energy-integrated angular profiles can be
nearly degenerate, while the energy spectra and energy-resolved angular
distributions remain distinct. 
Although the predicted signals may lie beyond the reach
of current observatories, these complementary spectral
and angular signatures provide potential targets for
future high-energy neutrino and gamma-ray searches.

\end{abstract}

\newpage

\section{Introduction}
\label{sec:intro}

Black holes are remarkable sites where gravity, quantum theory, and particle physics intersect. Hawking radiation
\cite{Hawking:1974rv,Hawking:1975vcx}, one of their central quantum
predictions, nevertheless remains unobserved. Sufficiently light black
holes, if they exist, evaporate by emitting all particle species that are kinematically
accessible at their Hawking temperature, making microscopic black holes
natural probes of physics beyond the Standard Model (SM). Detecting their
evaporation products would therefore open a direct observational window
onto black hole quantum physics at microscopic scales.

Microscopic black holes may form inside celestial objects through the accumulation and gravitational collapse of dark matter. Neutron stars provide particularly favorable environments for this process owing to their exceptionally deep gravitational potentials and high matter densities.  Dark matter
(DM) particles from the Galactic halo can be gravitationally captured and
accumulate in their interiors
(see, e.g., Refs.~\cite{Dimopoulos:1982cz,Kolb:1982si} for pioneering
works and Ref.~\cite{Bramante:2023djs} for a recent review on the topic).  Depending on their
mass, spin statistics, and self-interactions, the accumulated dark matter may
eventually become self-gravitating and undergo gravitational collapse,
forming a microscopic black hole
\cite{Goldman:1989nd,Gould:1989gw,Kouvaris:2010jy,Kouvaris:2011fi,
McDermott:2011jp,Kouvaris:2011gb, Kouvaris:2012dz, Kouvaris:2013kra, Bell:2013xk, Bramante:2013hn,
Bramante:2013nma}. 
If its initial mass is sufficiently small, Hawking evaporation dominates
over accretion, and the black hole evaporates before significant growth
can occur. 
Continued dark matter capture can then lead to repeated
episodes of microscopic black hole formation and evaporation \cite{Dalianis:2026jcl}. Such events
may occur in neutron stars throughout the local Galactic neighborhood, with
 enhanced rates expected in dark matter-rich environments such as the Galactic
Center \cite{Navarro:1995iw, Navarro:1996gj}.

The extremely dense stellar medium, however, presents an immediate
obstacle to observing the evaporation directly. Standard Model particles
produced by Hawking radiation are efficiently scattered and absorbed
inside the neutron star, depositing their energy in the stellar medium
and potentially contributing to neutron star heating
\cite{Saha:2025fgu}. A qualitatively different possibility arises if a
fraction of the Hawking luminosity is emitted into sufficiently weakly
interacting and long-lived particles beyond the Standard Model.  
Such
particles can traverse the neutron star without significant attenuation,
escape its surface, and subsequently decay into observable SM particles \cite{Dalianis:2026jcl}.
They thereby transport a fraction of the Hawking radiation out of the
otherwise opaque stellar environment. Related mechanisms, in which long-lived mediators escape an otherwise
opaque astrophysical body before decaying into observable particles, have
been extensively studied in the context of dark matter annihilation in the
Sun, Earth, and other celestial bodies
\cite{Pospelov:2008jd,Batell:2009zp,Bell:2011sn,Feng:2015hja,Feng:2016ijc,Kouvaris:2016ltf,
Allahverdi:2016fvl,Leane:2017vag,
Leane:2021ihh, Bose:2021yhz,Nguyen:2022zwb,Acevedo:2024ttq,
Bose:2024wsh}.

In this work, we investigate the observable signatures
of this hypothetical mechanism, with particular
emphasis on the angular morphology of the resulting emission.
 We denote the
long-lived particle generically by $S$ and formulate its propagation and
decay into a daughter particle $Y$. Depending on the underlying
interaction, the decay products may include photons, neutrinos, charged
leptons, or hadrons, with hadronic final states also generating secondary
photons and neutrinos through fragmentation and decay; see, e.g.,
Refs.~\cite{Page:1976wx,Carr:1976zz,
MacGibbon:1990zk,MacGibbon:1991tj, Ukwatta:2015iba, Carr:2020gox}.
We focus primarily on $Y=\gamma$ and $\nu$, which provide complementary
high-energy messengers. Charged particles may provide an additional
cosmic ray signature, see e.g. \cite{Feng:2016ijc,Kouvaris:2016ltf}, although their directional information depends on
their energy and charge owing to deflection by Galactic magnetic fields;
we do not pursue this channel here.

High-energy particles from the evaporation of light
black holes have recently received renewed attention,
particularly in connection with ultra-high-energy neutrino
observations \cite{Zantedeschi:2024ram,Boccia:2025hpm}.
Unlike primordial black holes (PBH), which spend most of their
lifetimes at comparatively low Hawking temperatures and
reach the TeV--PeV regime only during their final stages
of evaporation
\cite{Dave:2019epr,Anchordoqui:2025xug,Klipfel:2025jql},
the black holes considered here are born hot.
Indeed, the requirement that Hawking evaporation dominate
over accretion restricts their masses to the microscopic
regime, corresponding to initial Hawking temperatures
typically at the TeV scale or above.
Continued dark matter capture and collapse may moreover
lead to successive evaporation events, providing a
sustained source of energetic particles $Y$.
This has motivated the possibility that dark matter-induced
black hole formation and evaporation in neutron stars may contribute
to the observed high-energy neutrino flux
\cite{Dalianis:2026jcl}.
Moreover, the dense stellar medium  absorbs most Standard Model
Hawking products, naturally suppressing the direct gamma-ray
and charged-particle signatures subject to stringent
constraints on evaporating PBHs
\cite{Page:1976wx,Carr:1976zz,Carr:2020gox,Cholis:2026wug}.
Related neutrino signatures from dark matter-induced
microscopic black holes formed inside the Sun and Earth
have also been investigated in Ref.~\cite{Acevedo:2020gro}.

In general, the expected signals are very faint.
 The emission from an individual neutron star, accounting for successive black hole evaporation
events,
is unlikely to be detectable from Earth with current
instruments. Achieving a neutrino flux comparable
to that observed by IceCube would require the source
to lie within a fraction of a parsec.
The cumulative contribution from the Galactic
neutron star population, however, can substantially
enhance the signal, potentially bringing it within
reach of current or future observatories
\cite{Dalianis:2026jcl}.
 IceCube has established an astrophysical
neutrino flux extending to PeV energies
\cite{IceCube:2013low,IceCube:2014stg}, while KM3NeT has recently
reported the ultra-high-energy event KM3-230213A
\cite{KM3Net:2016zxf,KM3NeT:2025npi}\footnote{ 
A subsequent IceCube point-source search placed constraints
on steady and transient neutrino emission from the direction
of the event \cite{Abbasi:2026mak}.}.
 At comparable energies, gamma-ray
observations by H.E.S.S. and LHAASO, together with CTAO, probe the
high-energy sky with substantially finer angular resolution
\cite{HESS:2006fka,LHAASO:2021ozi,CTAConsortium:2017dvg}.
These observations further motivate searches for unconventional
high-energy sources with characteristic spectral and angular signatures.

An important consequence of a finite mediator lifetime is that the
observable particles are not produced at the position of the neutron
star. The mediator propagates a macroscopic distance before decaying, so
that its relativistic decay products can reach the observer from
directions displaced from the line of sight to the parent star. An
intrinsically point-like source may therefore acquire an extended and
calculable angular profile on the sky. In this work, we derive the joint energy--angular
distribution of the decay products, incorporating mediator propagation,
its energy-dependent decay length, and relativistic decay kinematics.
For relativistic mediators, a simple characteristic scaling emerges: the
angular extent is controlled primarily by the combination
$c\tau_S/D$ of the mediator proper lifetime and source distance. The
boost enhancement of the mediator decay length is largely compensated by
the narrowing of the relativistic decay cone. The observed morphology can
therefore provide direct information about microscopic mediator
properties.

We demonstrate the general viability of this mechanism and illustrate its
phenomenology with five representative long-lived states:
gravitationally coupled scalars, axion-like particles (ALPs),
dark photons $A'$, dark-$Z$ bosons, and
heavy neutral leptons. These examples encompass qualitatively different
masses, couplings, lifetimes, and decay channels, and demonstrate that
mediators emitted in Hawking evaporation can escape the neutron star and
decay over astrophysically relevant distances. We discuss these
scenarios only to the extent required to establish the relevant
mediator parameter space and observable decay channels.
Notably, if several long-lived species carry Hawking-emitted energy out of the host
object, their different proper lifetimes may imprint multiple characteristic
angular scales on the observable signal, giving
rise to a multi-component angular morphology.

Beyond its overall angular extent, the signal contains correlated spectral
and angular information. The decay topology determines the relation
between the mediator energy and the energies and directions of its
daughters, while the Hawking spectrum supplies a broad distribution of
mediator energies. We contrast this case with monochromatic mediator
production from dark matter annihilation, which provides an alternative
dark matter-driven mechanism for producing high-energy particles \cite{Pospelov:2008jd,Batell:2009zp,Bell:2011sn,Feng:2015hja,Feng:2016ijc,Kouvaris:2016ltf,
Allahverdi:2016fvl,Leane:2017vag,
Leane:2021ihh, Bose:2021yhz,Nguyen:2022zwb,Acevedo:2024ttq,
Bose:2024wsh}.
Distinguishing between the two mechanisms could provide information on
both the underlying production process---black hole evaporation versus
DM annihilation---and the nature of the dark matter itself. Interestingly, the energy-integrated angular profiles of the two mechanisms
can be nearly degenerate, whereas their energy spectra provide a clear
means of distinguishing them. The energy-resolved angular distributions
offer an additional probe of the underlying production mechanism and
mediator kinematics.

The present work thus complements our previous study of the total luminosity and energy spectrum of high-energy particles from DM-induced microscopic
black holes by addressing the spatial morphology of the signal \cite{Dalianis:2026jcl}.
The same
underlying mechanism may appear as an unresolved source or as spatially
extended emission, depending on the mediator lifetime, source distance,
and instrumental angular resolution. Neutrino and gamma-ray observations
provide complementary probes: neutrinos propagate essentially
unattenuated over Galactic distances, while gamma-ray telescopes offer
substantially finer directional information and are therefore particularly
well suited to resolving the angular structure considered here. 
 The predicted signals are generally  faint and
their angular extent may be challenging to resolve with
current instruments. 
Nevertheless, future improvements in instrumental
sensitivity and angular resolution could open
promising opportunities for detecting and
characterizing this hypothetical signal.

The paper is organized as follows. In Sec.~\ref{sec:setup} we briefly
outline dark matter accumulation and microscopic-black hole formation inside neutron
stars and review the resulting Hawking emission. In Sec.~\ref{sec:Escape}
we discuss the production, escape, and decay of long-lived mediators and
introduce representative particle-physics realizations as benchmarks,
illustrating both the general viability of the mechanism and its model
dependence. In Sec.~\ref{sec:angular} we derive the joint energy--angular
distribution of the observable decay products and identify its
characteristic angular scale. In Sec.~\ref{sec:results} we present the
resulting energy and angular signatures, compare Hawking production with
dark matter annihilation, and discuss their observational implications. We summarize
our conclusions in Sec.~\ref{sec:conclusions}.

\section{Microscopic black hole evaporation in celestial objects}
\label{sec:setup}
We consider Hawking evaporation from microscopic black holes formed through dark matter accumulation and collapse inside celestial objects, focusing on neutron stars as the primary astrophysical environment.  We review the Hawking emission of weakly interacting long-lived
particles capable of escaping the dense stellar medium and derive the
secondary spectra produced by their decays. These results provide the
particle-production input for the beyond the Standard Model (BSM)  scenarios of Sec.~\ref{sec:Escape}
and the energy--angular analysis of Sec.~\ref{sec:angular}.

\subsection{Black hole formation and evaporation}
\label{sec:BHformation}

In neutron stars, halo dark matter particles can become gravitationally
captured after losing energy through scattering with the stellar
constituents. As the captured population grows, it may, under suitable
conditions, become gravitationally unstable and collapse into a microscopic
black hole near the stellar center. The conditions for collapse depend on
the dark matter mass, spin statistics, self-interactions, and interactions
with stellar matter, and have been extensively studied in the
literature~\cite{Goldman:1989nd,Gould:1989gw,Kouvaris:2010jy,Kouvaris:2011fi,
McDermott:2011jp,Kouvaris:2011gb,Bell:2013xk,Bramante:2013hn,
Bramante:2013nma, Giffin:2021kgb, Adarsha:2026tpb, Sarkar:2026fge}. 

We take the formation of a microscopic Schwarzschild black hole near the
neutron star center as the starting point of our analysis. Its initial mass
$M_{\rm BH}$ determines the initial Hawking temperature,
\begin{equation}
 T_{\rm BH}
 =
 \frac{\hbar c^3}{8\pi G M_{\rm BH}}
 \simeq
 1~{\rm PeV}
 \left(
 \frac{1.06\times10^4~{\rm kg}}{M_{\rm BH}}
 \right).
 \label{eq:THawking}
\end{equation}
We focus on $M_{\rm BH}\lesssim10^6~{\rm kg}$, corresponding 
to initial Hawking temperatures in the TeV range or above,
and hence to emission at the energies relevant for the
signals considered below.
For the black hole masses considered here, Hawking
evaporation is assumed to dominate over accretion.
The precise condition depends on the accretion rates
of both baryonic matter and dark matter.
The black holes therefore evaporate promptly rather
than growing and ultimately destroying the neutron star.
 We assume that
such black holes are repeatedly formed and evaporate near the stellar
center, producing a sequence of short evaporation bursts~\cite{Dalianis:2026jcl}.
More massive black holes are colder and longer-lived and may instead enter
an accretion-dominated regime, eventually destroying the neutron
star.

For the masses of interest here, Hawking evaporation is extremely rapid,
with characteristic lifetime
\begin{equation}
 t_{\rm evap}
 \simeq
 4.1\times10^{-7}~{\rm s}
 \left(
 \frac{M_{\rm BH}}{10^4~{\rm kg}}
 \right)^3.
 \label{eq:tevap}
\end{equation}
An individual evaporation event is therefore effectively instantaneous on
astrophysical timescales and can be treated as a burst originating near the
neutron star center. 
As illustrated in the left panel of
Fig.~\ref{fig:spectra1}, the instantaneous spectrum
hardens rapidly as the black hole temperature increases
toward the end of evaporation.
Although resolving this temporal evolution from an
individual neutron star would be challenging,
repeated dark matter accumulation and collapse
could give rise to a characteristic sequence of
evaporation bursts.

The neutron star does not appreciably modify the local near-horizon
Hawking emission. For $M_{\rm BH}\sim10^4~{\rm kg}$, the gravitational
radius, $r_g=GM_{\rm BH}/c^2\sim10^{-23}~{\rm m}$, is negligible compared
with the characteristic length scales of the neutron star background,
$R_{\rm NS}\sim10~{\rm km}$. On the microscopic scale of the black hole,
the stellar background can therefore be regarded as locally flat, while
the black hole geometry itself is well approximated by an isolated
Schwarzschild spacetime. The corresponding greybody factors may thus be
taken to be those of an isolated Schwarzschild black hole. The neutron star
gravitational field instead affects the escaping radiation on macroscopic
scales, most notably through gravitational redshift.
Corrections to the Hawking evaporation rate due to Pauli blocking in the
degenerate neutron star medium have been discussed in
Ref.~\cite{Autzen:2014tza}.

The stellar gravitational potential nevertheless redshifts the emitted
particles. For a static, spherically symmetric spacetime,
\begin{equation}
 E_\infty=\sqrt{-g_{tt}(r)}\,E_{\rm loc}(r),
 \label{eq:energy-redshift-general}
\end{equation}
where $g_{tt}\to-1$ at infinity. In the constant-density approximation, the
Schwarzschild interior solution~\cite{Misner:1973prb} gives
\begin{equation}
 \sqrt{-g_{tt}(0)}
 =
 \frac{3}{2}
 \sqrt{1-\frac{2GM_{\rm NS}}{R_{\rm NS}c^2}}
 -\frac{1}{2}
 \simeq0.69,
 \label{eq:redshift}
\end{equation}
for $M_{\rm NS}=1.5\,M_\odot$ and $R_{\rm NS}=12~{\rm km}$. Thus,
$E_\infty\simeq0.69\,E_{\rm em}$ for particles emitted near the center.
Since the precise factor depends on the neutron star mass, radius, and
equation of state, we absorb this order-unity correction into the benchmark
energy scales quoted below. In the subsequent propagation analysis,
$E_\infty$ is understood as the mediator energy, and therefore determines
its boost and decay length.

Standard Model particles produced through Hawking evaporation
are efficiently absorbed by the dense stellar medium,
a key feature that distinguishes this scenario from
PBH evaporation in vacuum.
  The observable signal considered
here instead arises from sufficiently weakly interacting, long-lived
particles $S$ that escape the neutron star and decay into detectable
secondaries. Their escape and decay conditions, together with representative
particle-physics realizations, are discussed in Sec.~\ref{sec:Escape}.
We first determine the Hawking spectrum of the emitted mediators.

\subsection{Hawking emission and mediator spectrum}
\label{sec:spectra}

\begin{figure*}[t]
    \centering
    \includegraphics[width=0.48\textwidth]{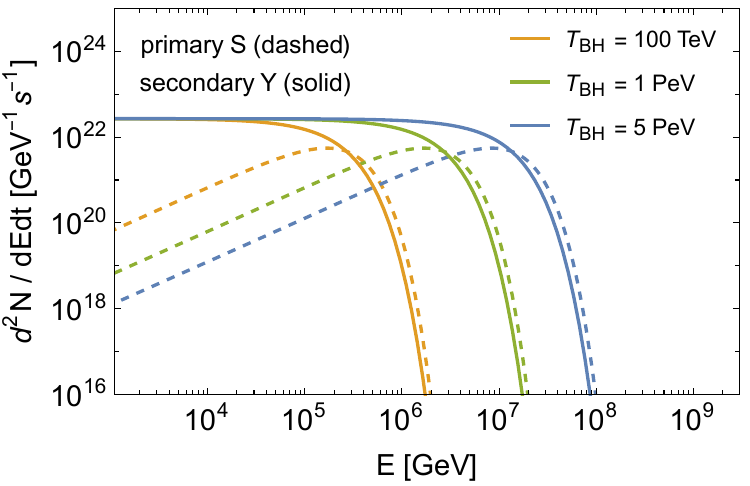} \,\,\,
    \includegraphics[width=0.48\textwidth]{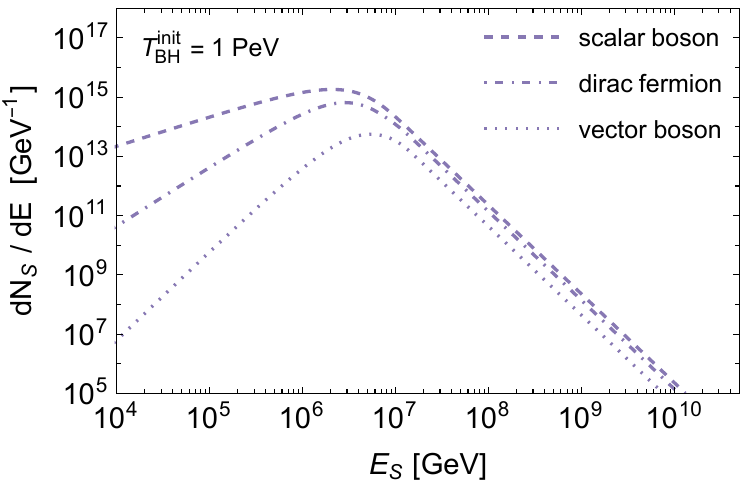}
 \caption{
{\small
{\it Left panel:} Instantaneous energy spectra of the
primary scalar mediator $S$ (dashed curves) and
secondary particles $Y$ (solid curves) for Hawking
temperatures $T_{\rm BH}=100~{\rm TeV}$, $1~{\rm PeV}$,
and $5~{\rm PeV}$.
The secondary spectra arise from the two-body decay
$S\to Y\bar{Y}$.
{\it Right panel:} Time-integrated Hawking energy spectra
of scalar, Dirac fermion, and vector mediators for an
initial Hawking temperature $T_{\rm BH}^{\rm init}=1~{\rm PeV}$,
accounting for the corresponding spin-dependent
greybody factors.
}}

\label{fig:spectra1}
\end{figure*}

A black hole radiates into all kinematically accessible particle species.
The instantaneous Hawking emission rate of a species $i$ with spin $s_i$
and $g_i$ internal degrees of freedom is
\begin{equation}
 \frac{d^2N_i}{dt\,dE}
 =
 \frac{g_i}{2\pi\hbar}\,
 \frac{{\cal G}_i(M_{\rm BH},E,m_i,s_i)}
 {e^{E/T_{\rm BH}}-(-1)^{2s_i}},
 \label{eq:hawking-primary}
\end{equation}
with the corresponding primary mediator spectra shown by the dashed curves
in the left panel of Fig.~\ref{fig:spectra1}. Here, ${\cal G}_i$ is the
dimensionless greybody factor, summed over angular-momentum modes. It
accounts for transmission through the black hole curvature potential and
introduces a spin- and energy-dependent modification of the thermal
spectrum.

During evaporation, the black hole mass decreases while its temperature
increases. The mass-loss rate can be parametrized as
\begin{equation}
 \frac{dM_{\rm BH}}{dt}
 =
 -\frac{\alpha(M_{\rm BH})}{M_{\rm BH}^2},
 \label{eq:BH-mass-loss}
\end{equation}
where $\alpha(M_{\rm BH})$ accounts for the particle degrees of freedom
accessible at the instantaneous Hawking temperature. The bulk of the
black hole mass is emitted while the Hawking temperature remains within
an order of magnitude of its initial value. 

The spectrum from a
single evaporation event is 
\begin{equation}
 \frac{dN_i}{dE}
 =
 \int_0^{t_{\rm evap}}dt\,
 \frac{d^2N_i}{dt\,dE}.
 \label{eq:time-integrated-spectrum}
\end{equation}
The asymptotic behavior of this spectrum follows directly from the scaling
of Hawking emission. Adopting natural units, $\hbar=c=1$, the greybody
factor for a massive particle depends on $GM_{\rm BH}E$ and $m_i/E$.
In the relativistic regime, $E\gg m_i$, it approaches its massless form
and can be written in terms of
\begin{equation}
 x\equiv\frac{E}{T_{\rm BH}}
 =8\pi GM_{\rm BH}E,
 \qquad
 {\cal G}_i(M_{\rm BH},E,m_i,s_i)
 \simeq{\cal G}_i(x,s_i).
 \label{eq:greybody-scaling}
\end{equation}

Approximating $\alpha(M_{\rm BH})$ as constant over the relevant stages of
evaporation, Eqs.~\eqref{eq:BH-mass-loss} and
\eqref{eq:time-integrated-spectrum} give
\begin{equation}
 \frac{dN_i}{dE}
 =
 \frac{g_i}{2\pi\alpha(8\pi)^3G^3}\,
 E^{-3}
 \int_{0}^{E/T_{\rm BH}^{\rm init}} dx\,
 \frac{x^2{\cal G}_i(x,s_i)}
 {e^x-(-1)^{2s_i}},
 \label{eq:BH-xint}
\end{equation}
where $T_{\rm BH}^{\rm init}$ is the initial Hawking temperature. For
$E\gg T_{\rm BH}^{\rm init}$, the integral approaches a constant, yielding the
universal high-energy behavior $dN_i/dE\propto E^{-3}$.

Below the initial Hawking temperature, the spectrum instead reflects the
spin-dependent low-frequency behavior of the greybody factors. For
$m_i\ll E\ll T_{\rm BH}^{\rm init}$,
\begin{equation}
 \frac{dN_i}{dE}\propto
 \begin{cases}
 E\,\,T{_{\rm BH}^{\rm init}}^{-4},   & s_i=0,\\
 E^2\,T{_{\rm BH}^{\rm init}}^{-5},   & s_i=\frac12,\\
 E^3\,T{_{\rm BH}^{\rm init}}^{-6},   & s_i=1,
 \end{cases}
 \label{eq:low-energy-scaling}
\end{equation}
while the transition around $E\sim T_{\rm BH}^{\rm init}$ depends on the full
greybody factor. 
These regimes are illustrated in the right panel of
Fig.~\ref{fig:spectra1}.
More detailed calculations of Hawking emission spectra can be
performed using dedicated numerical codes such as
\texttt{BlackHawk}~\cite{Arbey:2021mbl} or 
Ref. ~\cite{Carlini:2025bki} dedicated to gamma-ray emission.

We are particularly interested in a long-lived mediator $S$, which may be
a scalar, a Dirac or Majorana fermion, or a vector, emitted directly through
Hawking radiation. For the parameter space considered here,
$T_{\rm BH}^{\rm init}\gtrsim~{\rm TeV}$ and $m_S < T_{\rm BH}^{\rm init}$, so
its spectrum follows the asymptotic behaviors above, with the full greybody
spectrum retained around $E_S\sim T_{\rm BH}^{\rm init}$. This primary spectrum,
$dN_S/dE_S$, determines the secondary neutrino or photon spectrum after
the mediator escapes the star and decays. We next consider the two-body
decay $S\to Y\bar Y$, with $Y=\nu$ or $\gamma$.

\subsection{Secondary neutrino and photon spectra}
\label{sec:secondary}

The observable signal arises from mediator decays outside the neutron star.
We denote by $Y$ a stable daughter particle and focus on neutrinos and
photons, $Y=\nu,\gamma$, whose masses are negligible at the energies of
interest. We begin with, and primarily focus on, the two-body decay
\begin{equation}
 S\to Y\bar Y ,
 \label{eq:SYY}
\end{equation}
for which the secondary spectrum can be obtained analytically. In the mediator rest frame, denoted by an asterisk ($*$), the daughters
are emitted back to back, each with energy $E_Y^*=m_S/2$. For an isotropic two-body decay, the probability for either
daughter to be emitted into the solid-angle element $d\Omega^*$ is $dP={d\Omega^*}/{4\pi}$, or equivalently
\begin{equation} \label{eq:solidanglestar}
 \frac{dP}{d\Omega^*}=\frac{1}{4\pi}.
\end{equation}
In the stellar frame, where the mediator has energy $E_S$, Lorentz factor
$\gamma_S=E_S/m_S$, and velocity
$\beta_S=(1-m_S^2/E_S^2)^{1/2}$, the daughter energy is
\begin{equation}
 E_Y
 =
 \gamma_S E_Y^*
 \left(1+\beta_S\cos\theta^*\right)
 =
 \frac{E_S}{2}
 \left(1+\beta_S\cos\theta^*\right),
 \label{eq:EY-boost}
\end{equation}
where $\theta^*$ is the daughter emission angle relative to the mediator
direction of motion.

Since $\cos\theta^*$ is uniformly distributed in $[-1,1]$, the daughter
spectrum from a single mediator decay is a flat box
\cite{Ibarra:2012dw},
\begin{equation}
 \frac{dn_{S\to Y}(E_S,E_Y)}{dE_Y}
 =
 \frac{n_Y}{\beta_S E_S}\,
 \Theta(E_Y-E_Y^-)\,
 \Theta(E_Y^+-E_Y),
 \label{eq:box}
\end{equation}
with endpoints
\begin{equation}
 E_Y^\pm=\frac{E_S}{2}(1\pm\beta_S),
 \label{eq:box-endpoints}
\end{equation}
where $n_Y=2$ is the daughter multiplicity for
the two-body decay Eq. \eqref{eq:SYY}.

The secondary spectrum from a complete evaporation event follows by
convolving the mediator spectrum with the two-body decay kernel,
\begin{equation}
 \frac{dN_Y}{dE_Y}
 =
 {\rm Br}(S\to Y\bar Y)
 \int_{E_{S,\min}(E_Y)}^\infty dE_S\,
 \frac{dN_S}{dE_S}
 \frac{n_Y}{\beta_S E_S},
 \label{eq:secondary}
\end{equation}
where
\begin{equation}
 E_{S,\min}(E_Y)
 =
 E_Y+\frac{m_S^2}{4E_Y}.
 \label{eq:ESmin}
\end{equation}

For $\beta_S\ll1$, the box becomes narrow around
$E_Y\simeq m_S/2$, whereas in the ultrarelativistic limit,
$\beta_S\simeq1$, it extends approximately over
$0<E_Y<E_S$, and Eq.~\eqref{eq:box} reduces to
$(2/E_S)\Theta(E_S-E_Y)$.
Thus, a high-energy mediator spectrum
$dN_S/dE_S\propto E_S^{-3}$ yields the same asymptotic scaling for the
secondary spectrum,
\begin{equation}
 \frac{dN_Y}{dE_Y}\propto E_Y^{-3}.
 \label{eq:secondary-powerlaw}
\end{equation}
The resulting primary mediator and secondary-particle instantaneous spectra are shown in
the left panel of Fig.~\ref{fig:spectra1} for representative Hawking
temperatures.

The above expressions apply directly to $S\to\nu\bar\nu$ and
$S\to\gamma\gamma$, differing only through their branching fractions.
Charged final states may also provide observable cosmic-ray signatures,
although their directional information can be affected by Galactic magnetic
fields, depending on their energy and charge. More general decays, including
three-body modes and decay chains through intermediate particles, require
the corresponding decay kernels and may generate secondary neutrinos,
photons, and charged particles. Examples of direct and cascade production
are discussed in Sec.~\ref{sec:Escape}.

So far we have integrated over the emission directions. For relativistic
mediators, however, the daughter energy and direction are correlated by
the Lorentz boost. Together with the spatial distribution of mediator
decays, this correlation determines the angular morphology developed in
Sec.~\ref{sec:angular}. We now turn to representative BSM scenarios involving long-lived mediators.

%
%
%

\section{Escaping long-lived BSM mediators}
\label{sec:Escape}

An observable signal from black hole evaporation inside a neutron star
requires a Hawking-emitted, beyond the Standard Model, species $S$ to escape the dense stellar medium
and subsequently decay into observable Standard Model particles. Its
differential Hawking emission rate is given by Eq.~\eqref{eq:hawking-primary},
with $i=S$.

\subsection{Escape and decay requirements}

For a mediator produced inside a dense celestial object, escape without
significant interactions requires its scattering optical depth through the
surrounding matter to satisfy
\begin{equation}
 \tau_{\rm scat}(E_S)
 \simeq
 n_{\rm mat}\,\sigma_{S{\rm -mat}}(E_S)\,R
 \lesssim 1,
 \qquad
 P_{\rm esc}(E_S)\simeq e^{-\tau_{\rm scat}(E_S)},
 \label{eq:escape-condition}
\end{equation}
where $n_{\rm mat}$ is the characteristic number density of scattering
targets and $R$ the characteristic size of the object. For the neutron
stars considered here, $n_{\rm mat}\sim10^{38}\,{\rm cm^{-3}}$ and
$R=R_{\rm NS}\sim10~{\rm km}$. Equivalently, the mediator mean free path,
$\lambda_S=[n_{\rm mat}\sigma_{S{\rm -mat}}]^{-1}$, must satisfy
$\lambda_S\gtrsim R$. The same formalism applies to other celestial
objects upon replacing the appropriate density, size, and matter composition.

The mediator must also survive long enough to escape the star. Its boosted
decay length is
\begin{equation}
 \ell_{\rm dec}(E_S)
 =
 \gamma_S\beta_S c\tau_S
 =
 \gamma_S\beta_S\frac{\hbar c}{\Gamma_S},
 \label{eq:general-decay-length}
\end{equation}
so that
\begin{equation}
 \ell_{\rm dec}(E_S)\gtrsim R_{\rm NS},
 \label{eq:outside-condition}
\end{equation}
or, for relativistic mediators,
$\gamma_S\tau_S\gtrsim3.3\times10^{-5}\,{\rm s}$.

Conversely, the mediator must decay before reaching the observer to
produce the secondary signal considered here. For a source at distance
$D$, the probability of decay between the neutron star surface and the
observer is
\begin{equation}
 P_{\rm dec}(E_S;R_{\rm NS},D)
 =
 \exp\left[-\frac{R_{\rm NS}}{\ell_{\rm dec}(E_S)}\right]
 -
 \exp\left[-\frac{D}{\ell_{\rm dec}(E_S)}\right].
 \label{eq:Pdec_window}
\end{equation}
This expression continuously accounts for the fraction of mediators
decaying within the observable region.

Long-lived particles may additionally be subject to cosmological
constraints if populated in the early Universe. In particular, decays
during or after Big Bang Nucleosynthesis can modify light-element
abundances through electromagnetic or hadronic energy injection. Such
constraints depend on the primordial abundance, lifetime, and decay
channels, and are therefore model dependent.

We next consider, and briefly review,  representative realizations involving gravitationally
coupled particles, axion-like particles,  dark vector mediators, and heavy neutral leptons. These
examples illustrate different ways in which a Hawking-emitted state can
escape the star and convert the otherwise hidden evaporation into
an observable neutrino or photon signal.

\subsection{Gravitationally coupled long-lived particles} 
\label{sec:grav-mediator}

A simple realization of long-lived mediators is provided by
particles with Planck-suppressed interactions, such as moduli
and dilaton-like fields arising in string compactifications
\cite{Becker:2006dvp,Kiritsis:2019npv,Cicoli:2023opf},
or weakly coupled states associated with hidden and
supersymmetry-breaking sectors~\cite{Binetruy:2006ad}.
We denote such a particle by $X$ and assume that it is
emitted directly through Hawking radiation.

We define the Planck and reduced Planck masses as
\begin{equation}
 m_{\rm Pl}=G^{-1/2}\simeq1.22\times10^{19}~{\rm GeV},
 \qquad
 M_{\rm Pl}=(8\pi G)^{-1/2}\simeq2.44\times10^{18}~{\rm GeV}.
\end{equation}

Gravitational-strength interactions render the neutron star
effectively transparent to these particles. In particular,
the cross section for hard scattering on a neutron scales as
\begin{equation}
 \sigma_{X{\rm -mat}}(E_X)
 \sim \frac{s}{M_{\rm Pl}^4}
 \sim \frac{2m_nE_X}{M_{\rm Pl}^4},
 \label{eq:sigma_grav_est}
\end{equation}
up to factors of order unity, where
$m_n\simeq0.939~{\rm GeV}$. Even at $E_X\sim1~{\rm PeV}$,
the corresponding optical depth satisfies
$\tau_{\rm scat}\ll1$.

The decay width of a gravitationally coupled state
typically scales as
\begin{equation}
 \Gamma_X
 \simeq
 \frac{\tilde c}{16\pi}\frac{m_X^3}{M_{\rm Pl}^2},
 \label{eq:Gamma_modulus_scaling}
\end{equation}
where $\tilde c$ parametrizes the model-dependent
couplings and available decay channels
\cite{Nakamura:2006uc,Endo:2006zj,Cicoli:2015ylx}.
For $\beta_X\simeq1$, the boosted decay length is
\begin{equation}
 \ell_{\rm dec}(E_X)
 \simeq
 \frac{16\pi\hbar c}{\tilde c}
 \frac{M_{\rm Pl}^2E_X}{m_X^4}
 \simeq
 \frac{1.9}{\tilde c}\,{\rm pc}
 \left(\frac{E_X}{1~{\rm PeV}}\right)
 \left(\frac{1~{\rm TeV}}{m_X}\right)^4.
 \label{eq:Ldec_grav}
\end{equation}

\begin{figure}[!tp]
    \centering
    \begin{subfigure}[t]{0.48\textwidth}
        \centering
        \includegraphics[width=\textwidth]{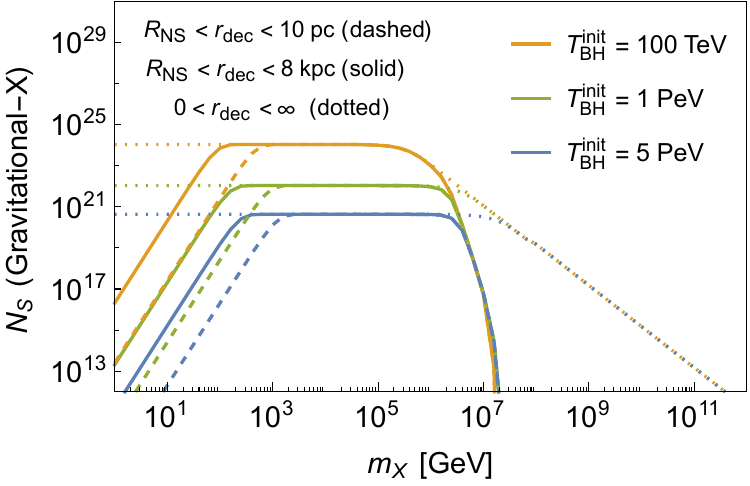}
    \end{subfigure}
    \hfill \,\,
    \begin{subfigure}[t]{0.48\textwidth}
        \centering
        \includegraphics[width=\textwidth]{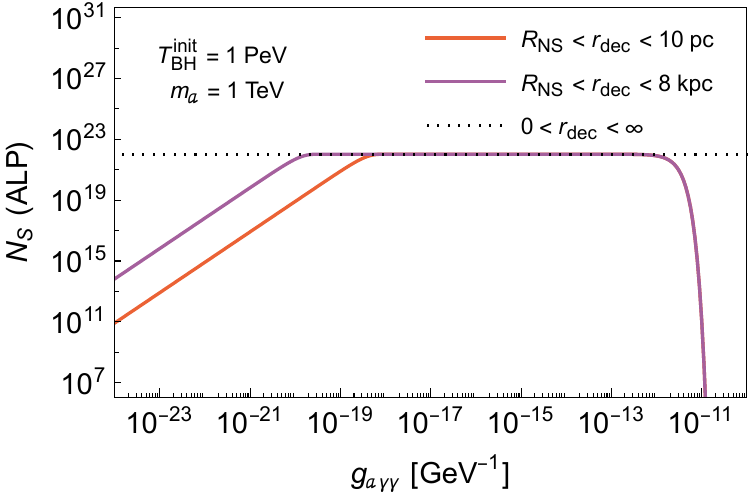}
    \end{subfigure}
  \par\vspace{0.4cm}
    \begin{subfigure}[t]{0.5\textwidth}
        \centering
        \includegraphics[width=\textwidth]{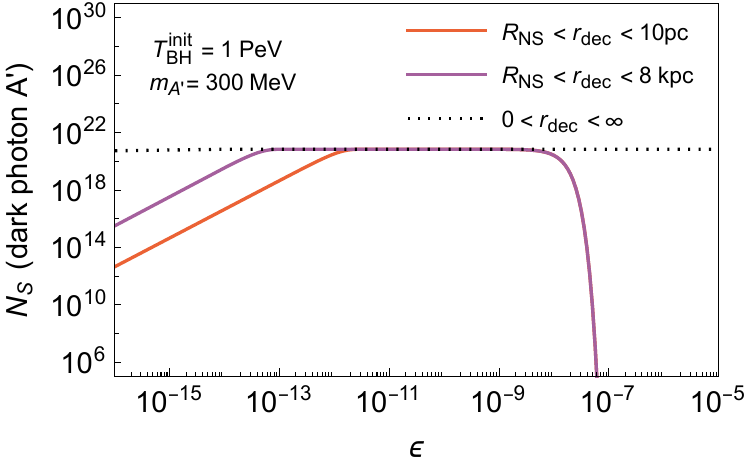}
    \end{subfigure}
    \hfill
    \begin{subfigure}[t]{0.48\textwidth}
        \centering
        \includegraphics[width=\textwidth]{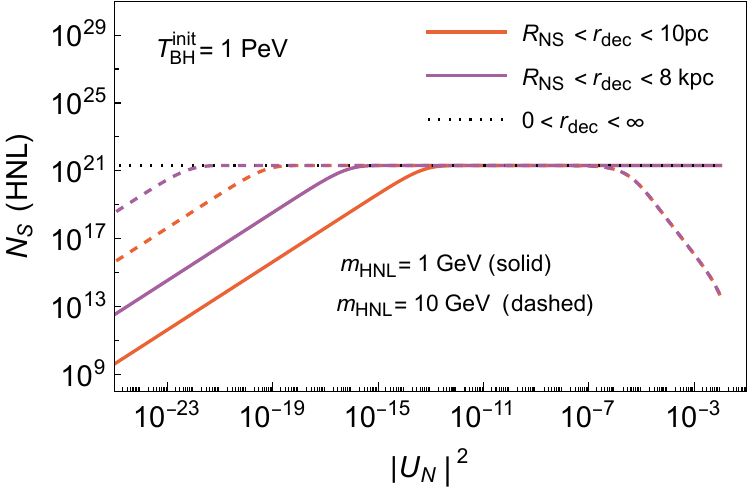}
    \end{subfigure}
\small{
\caption{
Number of long-lived BSM particles produced through Hawking
evaporation that escape the neutron star and decay within
the indicated radial intervals.
The four panels correspond to representative scenarios
discussed in Sec.~\ref{sec:Escape}.
{\it Upper left:} Gravitationally coupled scalar $X$
as a function of $m_X$, for
$T_{\rm BH}^{\rm init}=100~{\rm TeV}$, $1~{\rm PeV}$,
and $5~{\rm PeV}$.
{\it Upper right:} Axion-like particle (ALP) as a
function of the photon coupling $g_{a\gamma\gamma}$.
{\it Lower left:} Dark photon $A'$ as a function of
the kinetic mixing $\epsilon$, for
$m_{A'}=300~{\rm MeV}$ and $E_{A'}=1~{\rm PeV}$.
{\it Lower right:} Heavy neutral lepton $N$ as a
function of the active--sterile mixing $|U_N|^2$,
for $m_N=1$ and $10~{\rm GeV}$ and
$T_{\rm BH}^{\rm init}=1~{\rm PeV}$.
Solid and dashed curves represent particles escaping
the neutron star and decaying within the specified
distances, while dotted curves indicate the total
Hawking-emitted populations.
The suppression at large masses ($m_X$) or couplings
($g_{a\gamma\gamma}$, $\epsilon$, $|U_N|^2$) arises
from decays or attenuation within the neutron star,
whereas at small masses or couplings it reflects
particles decaying beyond the considered distances.
The resulting parameter windows identify regions of BSM parameter space
in which escaping mediators preferentially decay
within the relevant distance ranges, thereby
favoring potentially observable secondary neutrino
or gamma-ray signals.
}
\label{fig:NS_esc}
}
\end{figure}

The strong dependence $\ell_{\rm dec}\propto E_Xm_X^{-4}$
allows decay lengths ranging from the vicinity of the
neutron star surface to Galactic and larger scales.
For sufficiently light mediators, decays beyond the
Galactic scale suppress the Galactic signal and may
enhance the relative importance of extragalactic
contributions~\cite{Dalianis:2026jcl}.
The number of escaping mediators decaying within
representative distance ranges is shown in the upper
left panel of Fig.~\ref{fig:NS_esc}, while an illustrative
benchmark is summarized in Table~\ref{tab:LLP-benchmarks}.


The observable final states depend on the underlying
model. Direct decays,
\begin{equation}
 X\to\nu\bar\nu,\qquad X\to\gamma\gamma,
\end{equation}
provide examples of the two-body decay processes of
the scalar mediator $S$ considered in
Sec.~\ref{sec:angular}, while decays into charged
leptons, gauge bosons, or hadrons can produce
secondary neutrinos and photons through subsequent
decay chains and particle cascades.

\subsection{Axion-like particles}
\label{sec:ALP-mediator}

Another well-motivated class of long-lived mediators
consists of axion-like particles (ALPs), which arise
in extensions of the Standard Model involving broken
global symmetries or string compactifications
\cite{Arvanitaki:2009fg, Jaeckel:2010ni,Irastorza:2018dyq,DiLuzio:2020wdo}.
Unlike the QCD axion, a generic ALP can have
independently adjustable mass and couplings.
We denote the pseudoscalar mediator by $a$ and
assume that it is emitted directly through
Hawking radiation.

The ALP interaction with photons is described by
\begin{equation}
 \mathcal{L}_{a\gamma\gamma}
 =
 -\frac{1}{4}g_{a\gamma\gamma}
 a F_{\mu\nu}\widetilde F^{\mu\nu},
 \label{eq:ALP-interaction}
\end{equation}
where $g_{a\gamma\gamma}$ is a dimensionful coupling.
The corresponding two-photon decay width is
\begin{equation}
 \Gamma_{a\to\gamma\gamma}
 =
 \frac{g_{a\gamma\gamma}^2m_a^3}{64\pi},
 \label{eq:ALP-width}
\end{equation}
in natural units
\cite{Jaeckel:2010ni,Irastorza:2018dyq}.
Assuming that this channel dominates the total
decay width, the proper lifetime is
$\tau_a=64\pi\hbar/(g_{a\gamma\gamma}^2m_a^3)$.
For sufficiently weak photon couplings, ALPs can escape
the neutron star with negligible attenuation
($\tau_{\rm scat}\ll1$), provided their decay length
exceeds the stellar radius ($\ell_{\rm dec}\gtrsim R_{\rm NS}$).
Their propagation through the stellar medium
depends on the ALP energy, mass, and in-medium
electromagnetic interactions.

For relativistic ALPs, the boosted decay length is
\begin{align}
 \ell_{\rm dec}(E_a)
 &\simeq
 \frac{64\pi\hbar c\,E_a}
 {g_{a\gamma\gamma}^2m_a^4}
 \nonumber\\
 &\simeq
 0.13~{\rm pc}
 \left(\frac{E_a}{100~{\rm TeV}}\right)
 \left(\frac{1~{\rm GeV}}{m_a}\right)^4
 \left(\frac{10^{-12}~{\rm GeV}^{-1}}
 {g_{a\gamma\gamma}}\right)^2 .
 \label{eq:Ldec_ALP}
\end{align}
The strong dependence
$\ell_{\rm dec}\propto E_a g_{a\gamma\gamma}^{-2}m_a^{-4}$
allows a wide range of propagation distances.
For illustration, an ALP with $m_a=1~{\rm TeV}$,
$g_{a\gamma\gamma}=10^{-14}~{\rm GeV}^{-1}$,
and $E_a=1~{\rm PeV}$ has
$\tau_a\simeq1.3\times10^{-3}~{\rm s}$ and
$\ell_{\rm dec}\simeq4\times10^4 R_{\rm NS}$,
as summarized in Table~\ref{tab:LLP-benchmarks}.
The phenomenological viability of such benchmarks
depends on astrophysical and cosmological constraints
\cite{Cadamuro:2011fd,Depta:2020wmr}.

The direct decay
\begin{equation}
 a\to\gamma\gamma
\end{equation}
provides a simple realization of the two-body
kinematics considered in Sec.~\ref{sec:angular}.
Since the ALP is spinless, the photons are emitted
isotropically in its rest frame, and the
energy--angular distribution derived there
applies directly upon replacing $S$ by $a$.

\subsection{Dark vector mediators}
\label{sec:dark-vector}

Weakly coupled massive vector bosons provide another class
of long-lived mediators. We consider a dark photon $A'$,
interacting with the Standard Model through kinetic mixing
\cite{Pospelov:2008jd,Feng:2015hja,Feng:2016ijc,Kouvaris:2016ltf},
and a dark-$Z$ boson $Z'$ coupled to the SM neutral current.
Their different decay channels lead to distinct observable
signatures.

\subsubsection{Dark photon}
\label{sec:dark-photon}

The kinetically mixed dark photon is described by
\begin{equation}
 {\cal L}
 \supset
 -\frac{\epsilon}{2}F_{\mu\nu}F'^{\mu\nu}
 +\frac{1}{2}m_{A'}^2 A'_\mu A'^\mu ,
 \label{eq:dark-photon-lagrangian}
\end{equation}
where $\epsilon$ is the kinetic-mixing parameter.
We assume that $A'$ is emitted directly through
Hawking radiation as a massive spin-1 particle
with $g_{A'}=3$ degrees of freedom.

Interactions with the neutron star medium determine
the escape probability,
\begin{equation}
 \tau_{\rm tot}(E_{A'})
 =
 \tau_{\rm abs}(E_{A'})
 +
 \tau_{\rm scat}(E_{A'}),
 \qquad
 P_{\rm esc}(E_{A'})
 \simeq e^{-\tau_{\rm tot}(E_{A'})}.
 \label{eq:escape_general_Ap}
\end{equation}
Away from resonances, the mean free path scales as
$\lambda_{A'}\propto\epsilon^{-2}$
\cite{Hook:2021ous}. 

Outside the star, dark photons decay into
kinematically accessible charged SM particles.
For $m_{A'}>2m_\ell$, the leptonic partial width is
\begin{equation}
 \Gamma(A'\to\ell^+\ell^-)
 =
 \frac{\alpha_{\rm EM}}{3}\,
 \epsilon^2m_{A'}
 \left(1+\frac{2m_\ell^2}{m_{A'}^2}\right)
 \sqrt{1-\frac{4m_\ell^2}{m_{A'}^2}},
 \label{eq:Ap_ll}
\end{equation}
with additional hadronic contributions when
kinematically allowed~\cite{Ilten:2018crw}.
Since $\Gamma_{A'}\propto\epsilon^2m_{A'}$,
the boosted decay length scales as
\begin{equation}
 \ell_{\rm dec}(E_{A'})
 \simeq
 \frac{E_{A'}}{m_{A'}}
 \frac{\hbar c}{\Gamma_{A'}}
 \propto
 \frac{E_{A'}}{\epsilon^2m_{A'}^2}.
 \label{eq:Ldec_Ap}
\end{equation}
Thus, smaller $\epsilon$ facilitates escape but
increases the decay length, as illustrated in the
lower-left panel of Fig.~\ref{fig:NS_esc}.

For $m_{A'}=300~{\rm MeV}$,
$\epsilon=10^{-10}$, and $E_{A'}=1~{\rm PeV}$,
one obtains $\tau_{A'}\simeq4.8\times10^{-2}~{\rm s}$
and $\ell_{\rm dec}\simeq1.6\times10^{-3}~{\rm pc}$,
neglecting hadronic contributions, as summarized
in Table~\ref{tab:LLP-benchmarks}.

In the minimal kinetic-mixing scenario, $A'$ has
no tree-level coupling to neutrinos. Neutrinos
arise instead from secondary decays, such as
$A'\to\mu^+\mu^-$ followed by muon decay,
or from hadronic decay chains. These processes
can also produce high-energy photons through
radiation and electromagnetic or hadronic cascades,
requiring convolution over the secondary spectra.

\subsubsection{\texorpdfstring{Dark-$Z$ mediator}{Dark-Z mediator}}
\label{sec:dark-Z}

A dark-$Z$ boson couples to the SM neutral current,
\begin{equation}
 {\cal L}
 \supset
 g_{Z'} Z'_\mu J^\mu_{\rm NC},
 \label{eq:darkZ-lagrangian}
\end{equation}
where $g_{Z'}$ is a suppressed effective coupling.
Unlike the dark photon, $Z'$ couples directly to
neutrinos as well as charged fermions, with
model-dependent relative strengths
\cite{Davoudiasl:2012ag,Davoudiasl:2014kua}.

Its total decay width scales as
\begin{equation}
 \Gamma_{Z'}
 \sim C_{Z'}g_{Z'}^2m_{Z'},
 \label{eq:Gamma_Zp}
\end{equation}
where $C_{Z'}$ accounts for the accessible channels
and their electroweak couplings. The corresponding
boosted decay length is
\begin{equation}
 \ell_{\rm dec}(E_{Z'})
 \sim
 \frac{E_{Z'}}{m_{Z'}}
 \frac{\hbar c}{C_{Z'}g_{Z'}^2m_{Z'}}
 \propto
 \frac{E_{Z'}}{g_{Z'}^2m_{Z'}^2}.
 \label{eq:Ldec_Zp}
\end{equation}
Sufficiently weak couplings allow both free streaming
through the neutron star and macroscopic decay lengths.


The direct decay $Z'\to\nu\bar\nu$ provides
a concrete realization of the two-body decay
kinematics considered in Sec.~\ref{sec:angular}.
The decay width can be estimated as $ \Gamma_{Z'}\sim {\cal O}(1){g_{Z'}^2m_{Z'}}/{12\pi}$,
with the model dependence contained in the
order-unity factor. 
The corresponding decay properties for a representative
benchmark are summarized in
Table~\ref{tab:LLP-benchmarks}.

%
%

\subsection{Heavy neutral leptons}
\label{sec:HNL}

Heavy neutral leptons (HNLs) provide another well-motivated
class of long-lived particles
\cite{Atre:2009rg,Bondarenko:2018ptm},
whose phenomenology and experimental prospects have
been explored in numerous studies
\cite{Beltran:2021hpq,Beltran:2024twr}.
Here, we consider a Standard Model gauge-singlet
fermion $N$ that mixes weakly with active neutrinos,
as in sterile-neutrino extensions of the Standard Model,
and is emitted directly through Hawking radiation.
Related neutrino
signatures from long-lived sterile neutrinos produced
by dark matter annihilation in the Sun have been studied
in Ref.~\cite{Allahverdi:2016fvl}.
The relevant interactions are schematically
\begin{align}
 {\cal L}_{\rm int}
 \supset{}&
 -\frac{g}{\sqrt{2}}\,
 U_{\alpha N}W_\mu^+
 \overline{N}\gamma^\mu P_L\ell_\alpha
 -\frac{g}{2\cos\theta_W}\,
 U_{\alpha N}Z_\mu
 \overline{N}\gamma^\mu P_L\nu_\alpha
 +{\rm h.c.},
 \label{eq:HNL-interactions}
\end{align}
where $U_{\alpha N}$ denotes the active--sterile mixing,
with $\alpha=e,\mu,\tau$, and
$U_N^2\equiv\sum_\alpha|U_{\alpha N}|^2$.

The small mixing suppresses both HNL scattering
and decay. For momentum transfers below the
electroweak scale, the scattering cross section
scales as
\begin{equation}
 \sigma_{N{\rm -mat}}
 \sim U_N^2G_F^2s,
 \label{eq:HNL-scattering}
\end{equation}
up to process-dependent factors. At higher energies,
the full electroweak propagators must be included.
Sufficiently small mixing therefore allows HNLs
to escape the neutron star with negligible attenuation.

For $m_N\ll m_W$, decays proceed through off-shell
weak bosons, with total width
\begin{equation}
 \Gamma_N
 \sim C_N
 \frac{G_F^2m_N^5}{192\pi^3}U_N^2,
 \qquad
 \ell_{\rm dec}(E_N)
 \propto\frac{E_N}{U_N^2m_N^6},
 \label{eq:HNL-width-lowmass}
\end{equation}
where $C_N$ accounts for the accessible channels.
For $m_N\gtrsim m_W$, the two-body modes
\begin{equation}
 N\to\ell_\alpha^\mp W^\pm,\qquad
 N\to\nu_\alpha Z,\qquad
 N\to\nu_\alpha H
 \label{eq:HNL-two-body}
\end{equation}
become accessible. In the limit $m_N\gg m_W$,
\begin{equation}
 \Gamma_N
 \sim C_N'
 \frac{G_Fm_N^3}{8\sqrt{2}\pi}U_N^2,
 \qquad
 \ell_{\rm dec}(E_N)
 \propto\frac{E_N}{U_N^2m_N^4},
 \label{eq:HNL-width-highmass}
\end{equation}
where $C_N'$ depends on the decay channels and
the Dirac or Majorana nature of $N$.

Smaller $U_N^2$ facilitates escape but increases
the decay length, defining a parameter region in
which HNLs decay outside the neutron star yet
before reaching the observer. Their large Hawking
boost further extends the decay length.
Representative benchmarks are given in
Table~\ref{tab:LLP-benchmarks}, while the dependence
of the escaping and decaying population on $U_N^2$
is shown in the lower-right panel of
Fig.~\ref{fig:NS_esc}.

HNL decays produce both primary and secondary
neutrinos. Below the electroweak scale, relevant
channels include
\begin{equation}
 N\to\nu_\alpha\ell_\beta^+\ell_\beta^-,
 \qquad
 N\to\nu_\alpha\nu_\beta\bar\nu_\beta,
\end{equation}
as well as semileptonic and hadronic modes.
At higher masses, the two-body channels in
Eq.~\eqref{eq:HNL-two-body} produce neutrinos
directly or through subsequent SM decays.
The angular formalism of Sec.~\ref{sec:angular}
applies directly to primary neutrinos from
two-body decays, while three-body and cascade
channels require convolution over their
decay kinematics.

\begin{table*}[t]
\centering
\small
\renewcommand{\arraystretch}{1.25}
\setlength{\tabcolsep}{5pt}
\begin{tabular}{lcccccl}
\hline\hline
Mediator
& $m_S$
& Coupling
& $\gamma_S$
& $\tau_S$
& $\ell_{\rm dec}/R_{\rm NS}$
& Observable FS
\\
\hline

Gravitational $X$
& $100~{\rm TeV}$
& $\tilde c=1$
& $10$
& $0.20~{\rm s}$
& $6.0\times10^{4}$
& $\nu\bar\nu,\gamma\gamma$ (dir.)
\\

ALP $a$
& $1~{\rm TeV}$
& $g_{a\gamma\gamma}=10^{-14}~{\rm GeV}^{-1}$
& $10^{3}$
& $1.3\times10^{-3}~{\rm s}$
& $4.0\times10^{4}$
& $\gamma\gamma$ (dir.)
\\

Dark photon $A'$
& $300~{\rm MeV}$
& $\epsilon=10^{-10}$
& $3.3\times10^{6}$
& $4.8\times10^{-2}~{\rm s}$
& $4.9\times10^{9}$
& $\nu,\gamma$ (cas.)
\\

Dark-$Z$ $Z'$
& $10~{\rm GeV}$
& $g_{Z'}=5\times10^{-12}$
& $10^{5}$
& $0.10~{\rm s}$
& $3.1\times10^{8}$
& $\nu\bar\nu$ (dir.), $\gamma$ (cas.)
\\

HNL $N$
& $50~{\rm GeV}$
& $U_N^2=2\times10^{-19}$
& $2\times10^{4}$
& $0.5~{\rm s}$
& $3.1\times10^{8}$
& $\nu$ (dir./cas.), $\gamma$ (cas.)
\\

\hline\hline
\end{tabular}
\caption{
Representative benchmarks for long-lived particles emitted through
black hole evaporation inside a neutron star. Decay lengths are
evaluated at a reference mediator energy $E_S=1~{\rm PeV}$
and expressed in units of the neutron star radius,
$R_{\rm NS}=10~{\rm km}$.
The dark-$Z$ and HNL values are based on the approximate decay
widths discussed in the text and are intended as illustrative
benchmarks. ``Observable FS'' denotes the observable final
states, with ``dir.'' and ``cas.'' indicating direct and
cascade production, respectively.
}
\label{tab:LLP-benchmarks}
\end{table*}

%
%
%
%
%
%

\section{Angular Distribution of the Evaporation Signal}
\label{sec:angular}

Having reviewed representative BSM scenarios for the
mediator $S$, we now derive the joint energy--angular
distribution of secondary particles produced by
mediator decays outside the host celestial object.
The mediators are emitted with the Hawking spectrum derived in
Sec.~\ref{sec:spectra} and propagate and decay according to the framework
of Sec.~\ref{sec:Escape}. We denote the observable daughter particle
generically by $Y$, with the same geometry applying to neutrinos and
photons. While we focus on neutron stars, the formalism is readily
applicable to other celestial objects as well.

\subsection{Geometry and relativistic beaming}

The geometry and kinematics of displaced mediator decays have previously
been considered; see, e.g., Ref.~\cite{Kouvaris:2016ltf} for dark matter
annihilation in the Sun. Here we apply this framework to mediators produced
through Hawking evaporation and derive the resulting joint energy--angular
distribution.

The evaporation is assumed to occur at the stellar center. A mediator $S$
of energy $E_S$ and mass $m_S$ propagates radially outward and decays at a
distance $r$ from the star, located at distance $D$ from the Earth. We denote by $\theta$ the angle between the mediator momentum and the
Earth--star axis, by $\alpha$ the angle between the daughter and mediator
momenta in the stellar frame, and by $\psi$ the observed angular separation
between the daughter arrival direction and the position of the star on the
sky. Finally, $\theta^*$ denotes the daughter emission angle in the rest frame
of $S$, measured relative to the boost direction of the mediator, as introduced in
Sec.~\ref{sec:secondary}.

The geometry illustrated in Fig.~\ref{fig:triangle} gives
\begin{equation}
 r\sin\alpha=D\sin\psi,
 \label{eq:geometry}
\end{equation}
together with
\begin{equation}
 \alpha=\theta+\psi,
 \qquad
 D\cos\psi=r\cos\alpha+d,
 \label{eq:geometry-relations}
\end{equation}
where $d$ is the distance from the decay point to the Earth. For
$r\ll D$ and small observation angles,
\begin{equation}
 \alpha\simeq\frac{D}{r}\psi.
 \label{eq:alpha-small}
\end{equation}

Relativistic aberration relates the daughter direction in the mediator rest
frame to that in the stellar frame,
\begin{equation}
 \cos\alpha=
 \frac{\cos\theta^*+\beta_S}
      {1+\beta_S\cos\theta^*},
 \label{eq:aberration}
\end{equation}
where $\gamma_S=E_S/m_S$ and
$\beta_S=\sqrt{1-\gamma_S^{-2}}$ are the
mediator Lorentz factor and dimensionless velocity,
respectively.
For an isotropic two-body decay in the mediator rest frame,
$dP/d\Omega^*=1/(4\pi)$, as given in Eq.~\eqref{eq:solidanglestar}.
The corresponding angular distribution in the stellar frame is
\begin{equation}
 \frac{dP}{d\Omega_\alpha}
 =
 \frac{1}{4\pi\gamma_S^2}
 \frac{1}{(1-\beta_S\cos\alpha)^2}.
 \label{eq:beaming}
\end{equation}
In the ultra-relativistic, small-angle limit,
\begin{equation}
 \frac{dP}{d\Omega_\alpha}
 \simeq
 \frac{\gamma_S^2}
 {\pi\left(1+\gamma_S^2\alpha^2\right)^2},
 \label{eq:beaming-small}
\end{equation}
showing that the daughter emission is concentrated around the mediator
direction within a characteristic angle
\begin{equation}
 \alpha_{\rm beam}\sim\gamma_S^{-1}.
 \label{eq:beaming-angle}
\end{equation}
More quantitatively, approximately $50\%$ of the daughters are emitted
within $\alpha<\gamma_S^{-1}$, while about $99\%$ lie within
$\alpha<10\,\gamma_S^{-1}$.

Isotropic Hawking emission also produces mediators $S$ propagating away
from the observer. Their decay products can reach the Earth only if emitted
at sufficiently large angles relative to the mediator momentum, including
backward emission. Such configurations are strongly suppressed by
relativistic beaming for highly boosted mediators. We therefore neglect
this contribution in the forward, small-angle approximation adopted below.

\begin{figure}[!t]
\centering
\begin{tikzpicture}[scale=1.4]
\coordinate (S) at (0,4);
\coordinate (P) at (1,3);
\coordinate (B) at (0,0);
\node at (0,-0.2) {Earth};
\node at (0.4,4) {Star};
\node at (1.5,3.2) {$S\rightarrow Y\bar{Y}$};
\draw[thick,->] (S) -- (P) node[midway,right] {$r$};
\draw[dashed] (P) -- (2,2);
\draw[thick,->,blue] (P) -- (B) node[midway,left] {$Y$};
\node at (0.75,1.8) {$d$};
\draw[thick,->,blue] (P) -- (2,2.5) node[midway,above] {$\bar{Y}$};
\draw[dashed] (S) -- (B) node[midway,left] {$D$};
\fill (S) circle (1pt);
\fill (P) circle (1pt);
\fill (B) circle (1pt);
\draw (B)+(0,0.8) arc[start angle=90,end angle=72,radius=0.8];
\node at (0.15,0.95) {$\psi$};
\draw (S)+(0,-0.8) arc[start angle=-90,end angle=-46,radius=0.8];
\node at (0.25,3.1) {$\theta$};
\draw (P)+(-0.2,-0.5) arc[start angle=245,end angle=295,radius=0.8];
\node at (1.2,2.25) {$\alpha$};
\end{tikzpicture}
\caption{\small{ Geometry of the decay $S\rightarrow Y\bar{Y}$. The mediator is
emitted from the neutron star centre and decays at a distance $r$ from the
star and $d$ from the Earth, while the star lies at distance $D$. The angle
$\theta$ specifies the mediator direction relative to the Earth--star axis,
$\alpha$ is the angle between the observed daughter and the mediator
direction in the stellar frame, and $\psi$ is the angular separation between
the arrival direction of the daughter and the position of the star on the
sky.}}
\label{fig:triangle}
\end{figure}

\subsection{Energy--angle relation and decay kinematics}

For definiteness, consider the two-body decay
$S\rightarrow Y\bar Y$ into effectively massless daughters. In the rest
frame of $S$, each daughter has energy $m_S/2$, which transforms to the
stellar-frame energy given in Eq.~\eqref{eq:EY-boost}.
Using Eq.~\eqref{eq:aberration}, this can be written directly in terms of
the stellar-frame angle $\alpha$,
\begin{equation}
 E_Y=
 \frac{m_S^2}
 {2E_S(1-\beta_S\cos\alpha)}.
 \label{eq:EY-alpha-exact}
\end{equation}
For $\gamma_S\gg1$ and $\alpha\ll1$,
\begin{equation}
 E_Y\simeq
 \frac{E_S}{1+\gamma_S^2\alpha^2}
 \simeq
 \frac{E_S}
 {1+\displaystyle
 \frac{E_S^2D^2\psi^2}{m_S^2r^2}},
 \label{eq:EY-alpha}
\end{equation}
where Eq.~\eqref{eq:alpha-small} was used in the second expression. Thus, the daughter energy $E_Y$ and the observed angle $\psi$ are
intrinsically correlated.

For fixed $(E_Y,r,\psi)$, Eq.~\eqref{eq:EY-alpha} generally
admits two parent energies. Introducing the dimensionless discriminant
\begin{equation}
 \Delta(E_Y,r,\psi)
 =
 1-\frac{4E_Y^2D^2\psi^2}{m_S^2r^2}
 =
 1-\frac{r_{\rm min}^2}{r^2},
 \label{eq:Delta}
\end{equation}
the two solutions are
\begin{equation}
 E_S^\pm(E_Y,r,\psi)
 =
 \frac{m_S^2r^2}{2E_YD^2\psi^2}
 \left(1\pm\sqrt{\Delta}\right).
 \label{eq:Estar}
\end{equation}
Real solutions require $\Delta\geq0$, which imposes
\begin{equation}
 r\geq r_{\rm min}(E_Y,\psi)
 =
 \frac{2E_YD\psi}{m_S}.
 \label{eq:rmin}
\end{equation}
Thus, at fixed daughter energy, larger angular separations
require decays at progressively larger distances from the source. Since the decay radius is restricted to $r\leq D$, this also implies
 $\psi\leq{m_S}/{2E_Y}$,
which defines an energy-dependent kinematic upper bound on the
angular separation within the small-angle approximation.

The two branches have a simple physical interpretation. At fixed nonzero
$\alpha$, $E_Y$ initially increases with $E_S$, reaches a maximum near
$\gamma_S\alpha\simeq1$, and then decreases once the relativistic beaming
cone becomes narrower than $\alpha$. The same $(E_Y,\psi)$ at a given
decay radius can therefore originate from either the lower-energy branch
$E_S^-$ or the more highly boosted branch $E_S^+$.

The probability for a mediator of energy $E_S$ to decay between $r$ and
$r+dr$ is
\begin{equation}
 dP_{\rm dec}(E_S,r)
 =
 \frac{dr}{\ell_{\rm dec}(E_S)}
 \exp\left[-\frac{r}{\ell_{\rm dec}(E_S)}\right],
 \label{eq:decayprob-angular}
\end{equation}
with
$ \ell_{\rm dec}(E_S)=\gamma_S\beta_S c\tau_S$, Eq. \eqref{eq:general-decay-length}.
The two parent-energy branches are therefore weighted differently by both
the Hawking spectrum and their respective decay lengths,
$\ell_{\rm dec}(E_S^-)$ and $\ell_{\rm dec}(E_S^+)$.

\subsection{Differential fluence}
For isotropic Hawking emission, the number of mediators emitted into a
solid-angle element $d\Omega_S$ is
\begin{equation}
 dN_S=N_S\,\frac{d\Omega_S}{4\pi},
 \qquad
 d\Omega_S=2\pi\sin\theta\,d\theta.
 \label{eq:dNS-solid-angle}
\end{equation}
Consider a mediator that subsequently decays at a point located a distance
$d$ from the Earth. An area element $dA$ at the detector subtends the
solid angle $d\Omega_\alpha=dA/d^2$ at the decay point. The number of
daughter particles directed through this area is therefore
\begin{equation}
 dN_Y
 =
 n_Y\,{\rm Br}_{S\to Y}\,
 dN_S\,
 \frac{dP}{d\Omega_\alpha}\,
 \frac{dA}{d^2},
 \label{eq:dNY-schematic}
\end{equation}
where $n_Y$ is the daughter multiplicity per mediator decay. We use ${\rm Br}_{S\to Y}$ as a general shorthand for the branching
fraction into final states containing $Y$, while
${\rm Br}(S\to Y\bar Y)$ denotes the specific two-body channel considered
here.

We characterize the signal at Earth by the differential fluence
\begin{equation}
 \frac{d^2{\cal F}_Y}{dE_Y\,d\Omega_\psi}
 \equiv
 \frac{dN_Y}{dA\,dE_Y\,d\Omega_\psi},
 \label{eq:fluence-def}
\end{equation}
where $d\Omega_\psi=\sin\psi\,d\psi\,d\phi_\psi$. Before resolving the
daughter energy, the corresponding angular distribution is obtained by
expressing the mediator emission element $d\Omega_S$ in terms of the
observed solid-angle element $d\Omega_\psi$,
\begin{equation}
 \frac{dN_Y}{dA\,d\Omega_\psi}
 =
 \frac{n_Y\,{\rm Br}_{S\to Y}\,N_S}{4\pi}\,
 \frac{1}{d^2}
 \frac{d\Omega_S}{d\Omega_\psi}
 \frac{dP}{d\Omega_\alpha}.
 \label{eq:dNY-observed-angle}
\end{equation}

The geometrical mapping between the initial mediator direction and the
observed direction gives
\begin{equation}
 \frac{d\Omega_S}{d\Omega_\psi}
 =
 \frac{d^2}{r^2|\cos\alpha|}.
 \label{eq:angular-jacobian}
\end{equation}
For the forward branch, $\cos\alpha>0$, the geometrical dilution, the
angular Jacobian of Eq.~\eqref{eq:angular-jacobian}, and the daughter
beaming distribution combine to give
\begin{equation}
 \frac{1}{d^2}
 \frac{d\Omega_S}{d\Omega_\psi}
 \frac{dP}{d\Omega_\alpha}
 \simeq
 \frac{\gamma_S^2}
 {\pi r^2
 \left(1+\gamma_S^2D^2\psi^2/r^2\right)^2},
 \label{eq:geometric-beaming}
\end{equation}
where on the right-hand side we have used the ultra-relativistic,
small-angle approximations $\cos\alpha\simeq1$ and
$\alpha\simeq D\psi/r$.

To resolve the daughter energy, the angular decay factor
$n_Y\,dP/d\Omega_\alpha$ in Eq.~\eqref{eq:dNY-observed-angle} must be
promoted to the joint energy--angular decay kernel,
\begin{equation}
 \frac{d^2n_{S\to Y}}
      {dE_Y\,d\Omega_\alpha}
 =
 n_Y\,
 \frac{dP}{d\Omega_\alpha}\,
 \delta\!\left[
 E_Y-E_Y(E_S,\alpha)
 \right].
 \label{eq:joint-decay-kernel}
\end{equation}
Here $n_Y=2$ for $S\to Y\bar Y$, the angular distribution is given by
Eq.~\eqref{eq:beaming-small}, and $E_Y(E_S,\alpha)$ is the correlated
energy--angle relation of Eq.~\eqref{eq:EY-alpha}. Upon integration over
the daughter direction, Eq.~\eqref{eq:joint-decay-kernel} reproduces the
energy-only box spectrum of Eq.~\eqref{eq:box}.
Folding  the joint energy--angular kernel of Eq.~\eqref{eq:joint-decay-kernel} with the primary
Hawking spectrum and the mediator decay probability along its trajectory,
we obtain the observed joint energy--angular fluence
\begin{align}
 \frac{d^2{\cal F}_Y}{dE_Y\,d\Omega_\psi}
 ={}&
 \frac{n_Y\,{\rm Br}(S\to Y\bar Y)}{4\pi^2}
 \int_0^{t_{\rm evap}}dt
 \int dE_S\,
 \frac{d^2N_S}{dt\,dE_S}
 \int_{R_{\rm NS}}^{D}dr\,
 \frac{e^{-r/\ell_{\rm dec}(E_S)}}
      {\ell_{\rm dec}(E_S)}
 \nonumber\\
 &\times
 \frac{\gamma_S^2}
 {r^2
 \left(1+\gamma_S^2D^2\psi^2/r^2\right)^2}
 \,
 \delta\!\left[
 E_Y-
 \frac{E_S}
 {1+\gamma_S^2D^2\psi^2/r^2}
 \right].
 \label{eq:master-delta}
\end{align}
The factor $1/(4\pi^2)$ combines the $1/(4\pi)$ from isotropic Hawking
emission with the $1/\pi$ normalization of the relativistic daughter
beaming distribution. The delta function retains the correlation between
the daughter energy and arrival direction. 
Other decay
topologies can be incorporated by replacing the corresponding
multiplicity, branching fraction, and decay kernel.

The $E_S$ integral in Eq.~\eqref{eq:master-delta} can be performed
analytically using the two kinematic roots $E_S^\pm$ of
Eq.~\eqref{eq:Estar}, yielding
\begin{align}
 \frac{d^2{\cal F}_Y}
 {dE_Y\,d\Omega_\psi}
 ={}&
 \frac{n_Y\,{\rm Br}(S\to Y\bar Y)}{4\pi^2}
 \int_0^{t_{\rm evap}}dt
 \int_{\max(R_{\rm NS},r_{\rm min})}^{D}dr\,
 \frac{E_Y}
 {m_S^2r^2\,\sqrt{\Delta}}
 \nonumber\\
 &\times
 \sum_{i=\pm}
 E_S^i
 \left.
 \frac{d^2N_S}{dt\,dE_S}
 \frac{\exp[-r/\ell_{\rm dec}(E_S)]}
      {\ell_{\rm dec}(E_S)}
 \right|_{E_S=E_S^i}.
 \label{eq:master}
\end{align}
The lower radial limit incorporates the kinematic condition
$r\geq r_{\rm min}$ of Eq.~\eqref{eq:rmin}.
The dependence on the observation angle $\psi$ is implicit
through the discriminant $\Delta$, the parent energies $E_S^\pm$,
and the lower radial integration limit $r_{\rm min}$.
Thus, at fixed $E_Y$, the angular separation $\psi$ determines which
mediator energies and decay radii contribute to the observed signal.
For
$\ell_{\rm dec}\ll D$, the upper limit may be extended to infinity.
Although the factor $\Delta^{-1/2}$ diverges as $r\to r_{\rm min}$, the
resulting square-root singularity is integrable and does not lead to a divergence in the fluence.

The energy spectrum and angular profile are finally obtained by
integrating the joint distribution over the observed angle and daughter
energy, respectively,
\begin{align}
 \frac{d{\cal F}_Y}{dE_Y}
 &=
 \int d\Omega_\psi\,
 \frac{d^2{\cal F}_Y}{dE_Y\,d\Omega_\psi},
 &
 \frac{d{\cal F}_Y}{d\Omega_\psi}
 &=
 \int dE_Y\,
 \frac{d^2{\cal F}_Y}{dE_Y\,d\Omega_\psi}.
 \label{eq:integrated-fluence}
\end{align}
\subsection{Characteristic angular scale}

The characteristic angular extent of the signal has a simple physical
interpretation. A relativistic mediator typically propagates a distance
$r\sim\ell_{\rm dec}=\beta_S\gamma_S c\tau_S$ before decaying, while its
daughters are emitted within an angle $\alpha\sim\gamma_S^{-1}$ around
the mediator direction. Since $\psi\simeq(r/D)\alpha$ in the small-angle
limit,
\begin{equation}
 \psi_{\rm char}
 \sim
 \frac{\ell_{\rm dec}}{\gamma_S D}
 =
 \frac{\beta_S c\tau_S}{D}.
 \label{eq:psichar}
\end{equation}
Thus, for an ultra-relativistic mediator, the boost enhancement of the
decay length is compensated by the narrowing of the relativistic beaming
cone, leaving $\psi_{\rm char}\sim c\tau_S/D$.

This cancellation characterizes the typical angular scale, but does not
remove the energy dependence of the full energy-resolved morphology.
The joint distribution in Eq.~\eqref{eq:master} retains the correlation
between $E_Y$ and $\psi$, as well as the dependence on the primary Hawking
spectrum and mediator decay probability. The cancellation becomes
particularly transparent after integrating over the daughter energy.
From Eq.~\eqref{eq:master-delta}, one obtains
\begin{equation}
 \frac{d{\cal F}_Y}{d\Omega_\psi}
 =
 \frac{n_Y{\rm Br}_{S\to Y}}{4\pi^2}
 \int dE_S\,\frac{dN_S}{dE_S}
 \int_{R_{\rm NS}}^{D}dr\,
 \frac{e^{-r/\ell_{\rm dec}}}{\ell_{\rm dec}}
 \frac{\gamma_S^2}
 {r^2\left(1+\gamma_S^2D^2\psi^2/r^2\right)^2}.
 \label{eq:angular-spectrum}
\end{equation}
Here $dN_S/dE_S$ denotes the time-integrated mediator spectrum over the
complete black hole evaporation.

In the ultra-relativistic limit,
$\ell_{\rm dec}\simeq\gamma_S c\tau_S$. Introducing
\begin{equation}
 u\equiv\frac{r}{\ell_{\rm dec}},
 \qquad
 \xi\equiv\frac{D\psi}{c\tau_S},
 \label{eq:angular-variables}
\end{equation}
removes the explicit mediator-energy dependence from the radial kernel.
When
\begin{equation} \label{eq:elldec}
 R_{\rm NS}\ll\ell_{\rm dec}\ll D,
\end{equation}
the radial limits may be extended to $(0,\infty)$, yielding
\begin{equation}
 \frac{d{\cal F}_Y}{d\Omega_\psi}
 \simeq
 \frac{n_Y{\rm Br}_{S\to Y}\,N_S}
 {4\pi^2c^2\tau_S^2}\,
 {\cal A}(\xi),
 \qquad
 N_S\equiv\int dE_S\,\frac{dN_S}{dE_S},
 \label{eq:angular-factorized}
\end{equation}
where
\begin{equation}
 {\cal A}(\xi)
 =
 \int_0^\infty du\,
 \frac{e^{-u}u^2}{(u^2+\xi^2)^2}.
 \label{eq:Aangular}
\end{equation}
The Hawking spectrum therefore determines only the overall normalization
in this limit, while the normalized angular profile is universal and
depends only on the dimensionless angular variable $\xi$. If the hierarchy
in Eq.~\eqref{eq:elldec} is not satisfied, the finite radial boundaries
introduce a residual dependence on $\gamma_S$.

\begin{figure}[t]
    \centering
    \includegraphics[width=0.5\columnwidth]{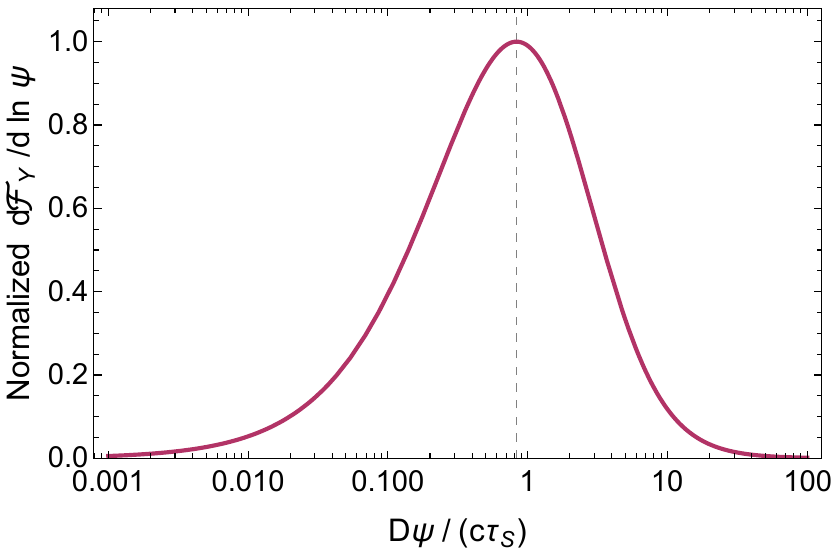}
    \caption{
    Normalized fluence per logarithmic angular annulus as a function
of the dimensionless angular variable
$\xi=D\psi/(c\tau_S)$, shown on a logarithmic scale.
The profile peaks at $\xi_{\rm peak}\simeq0.83$,
indicated by the vertical dashed line.
In the relativistic regime,
$R_{\rm NS}\ll\ell_{\rm dec}\ll D$,
the normalized angular profile is approximately independent
of the mediator energy, with its characteristic scale
determined by the mediator proper lifetime and source distance.
    }
    \label{fig:angular-log-fluence}
\end{figure}

The surface brightness $d{\cal F}_Y/d\Omega_\psi$ is maximal at
$\psi=0$. A more useful measure of the characteristic angular extent is
the fluence per logarithmic angular interval, or equivalently per
logarithmic annulus on the sky,
\begin{equation}
 \frac{d{\cal F}_Y}{d\ln\psi}
 \simeq
 2\pi\psi^2
 \frac{d{\cal F}_Y}{d\Omega_\psi}
 \propto
 \xi^2{\cal A}(\xi),
 \label{eq:fluence-logpsi}
\end{equation}
which peaks at $\xi\simeq0.83$. Hence
\begin{equation}
 \psi_{\rm peak}\simeq
 0.83\,\frac{c\tau_S}{D}.
 \label{eq:psipeak}
\end{equation}
The corresponding normalized angular profile is shown in
Fig.~\ref{fig:angular-log-fluence}. For a source at known distance, its angular extent therefore provides a
direct probe of the mediator proper lifetime. For example,
$D=10~{\rm pc}$ and $\tau_S=10^4~{\rm s}$ give
$\psi_{\rm peak}\simeq1.7~{\rm arcsec}$. Consequently, multiple long-lived
mediator species could imprint distinct angular scales according to their
different proper lifetimes.

This characteristic scale should be distinguished from the maximum angular separation allowed by the exact source--decay--observer geometry. For a decay at fixed
radius $r<D$, Eq.~\eqref{eq:geometry} gives
\begin{equation}
 \psi_{\rm max}^{\rm geom}(r)
 =
 \arcsin\left(\frac{r}{D}\right),
 \label{eq:psimaxgeom}
\end{equation}
reached at $\alpha=\pi/2$ (where the angular Jacobian of Eq.~\eqref{eq:angular-jacobian} develops an integrable square-root singularity). This geometric bound lies outside the
forward, small-angle regime used in deriving Eq.~\eqref{eq:master} and
therefore does not represent a sharp cutoff of the approximate angular
distribution. For relativistic mediators, the fluence is instead
concentrated at the much smaller characteristic scales of
Eqs.~\eqref{eq:psichar} and \eqref{eq:psipeak}.


\section{Observable Signatures of Black Hole Evaporation}
\label{sec:results}

Let us now discuss the observable signatures of the black hole evaporation
scenario developed above. We first estimate the characteristic luminosity
and event rates for the Galactic-center neutron star population and for a
single nearby neutron star. We then turn to the spectral and angular
morphology of the signal and examine to what extent these observables can
distinguish Hawking emission from alternative mechanisms of mediator
production.

\subsection{Luminosity and event rates}
\label{sec:event-rate}

Dark matter capture supplies the energy for the signal.
Following Ref.~\cite{Dalianis:2026jcl}, the capture power
of a benchmark neutron star with $M_{\rm NS}=1.4\,M_\odot$
and $R_{\rm NS}=10\,{\rm km}$ is approximately
\begin{equation}
\label{eq:NS-capture-power}
\dot E_{\rm cap}^{\rm NS}
\simeq
1.3\times10^{29}\,{\rm GeV\,s^{-1}}\,
\tilde{\sigma}_{\chi n}
\left(\frac{\rho_\chi}{10^3\,{\rm GeV\,cm^{-3}}}\right)
\left(\frac{220\,{\rm km\,s^{-1}}}{\bar{u}_\infty}\right),
\end{equation}
where $\rho_\chi$ is the ambient density of the captured
dark matter species, $\bar{u}_\infty$ is its characteristic
speed far from the star in the stellar rest frame, and
$\tilde{\sigma}_{\chi n}=1$ corresponds to geometrically
saturated capture.

Integrating Eq.~\eqref{eq:NS-capture-power} over the benchmark
neutron star population near the Galactic Center, assuming
saturated capture and an NFW dark matter profile, gives
\begin{equation}
\label{eq:GC-capture-power}
\dot E_{\rm cap}^{\rm GC}
\simeq 2.1\times10^{35}\,{\rm GeV\,s^{-1}}
\simeq 0.1\,L_\odot.
\end{equation}
A steeper inner dark matter profile can increase this power,
with the enhancement depending also on how strongly the neutron
star population is concentrated toward the center.

If the captured dark matter is ultimately converted into
Hawking radiation through repeated black hole formation
and evaporation, the time-averaged luminosity emitted in
mediators $S$ is
\begin{equation}
L_S \simeq f_{S|H}\,\dot E_{\rm cap},
\end{equation}
where $f_{S|H}$ is the fraction of Hawking-emitted energy
carried by $S$. The observable photon or neutrino flux also
depends on the mediator escape and decay probabilities, Eqs.~\eqref{eq:escape-condition} and \eqref{eq:Pdec_window},
the decay branching fractions, and the energy and angular
distributions of the daughter particles. 
For the NFW benchmark, assuming an optimistic spectral
efficiency for either
neutrinos or photons at the relevant energies,
the Galactic Center energy flux is estimated as
\begin{equation}
\label{eq:power_estimate}
E_Y^2\frac{d\Phi_Y^{\rm GC}}{dE_Y}
\sim 2.5\times10^{-13}\,
{\rm GeV\,cm^{-2}\,s^{-1}}\,.
\end{equation}

For a single nearby neutron star, the available energy flux
scales as
\begin{equation}
\label{eq:local-NS-scaling}
\frac{\dot E_{\rm cap}^{\rm NS}}{4\pi D^2}
\propto
\frac{\rho_\chi}{\bar{u}_\infty D^2}
\end{equation}
at fixed stellar properties and capture efficiency, where
$D$ is the distance to the star. For a local dark matter
density $\rho_{\chi\odot}\simeq0.4\,{\rm GeV\,cm^{-3}}$,
$\bar{u}_\infty=220\,{\rm km\,s^{-1}}$, and saturated
capture, Eq.~\eqref{eq:NS-capture-power} gives
$\dot E_{\rm cap}^{\rm local}\simeq
5.1\times10^{25}\,{\rm GeV\,s^{-1}}$.
The available energy fluxes from this star and the Galactic
Center population are equal at
\begin{equation}
\label{eq:local-GC-equivalence}
D_{\rm eq}
\simeq D_{\rm GC}
\left(
\frac{\dot E_{\rm cap}^{\rm local}}
     {\dot E_{\rm cap}^{\rm GC}}
\right)^{1/2}
\simeq 0.13\,{\rm pc},
\end{equation}
where $D_{\rm GC}\simeq8.3\,{\rm kpc}$. At
$D=120\,{\rm pc}$, representative of the nearest known
neutron stars, the local-to-Galactic-Center flux ratio
is only $\simeq1.2\times10^{-6}$. This ratio compares the
energy budgets before accounting for mediator propagation
and decay. In particular, long-lived mediators may decay
outside the region of interest, as illustrated in
Fig.~\ref{fig:NS_esc}.

The conversion of the predicted fluxes into detected events
depends strongly on the final-state particle and the detector
response. For neutrinos, the expected event numbers from the
Galactic-center population are small, as discussed in
Ref.~\cite{Dalianis:2026jcl}. The observable rate is obtained
by folding the differential neutrino flux with the
energy- and direction-dependent detector effective area,
which accounts for the interaction probability, detector
response, and event-selection efficiency. In particular,
the number of neutrinos geometrically crossing a detector-sized
area should not be identified with the number of detected events.

Gamma rays produced directly or through mediator decays may
offer complementary detection prospects. At multi-TeV energies,
ground-based gamma-ray observatories can achieve large effective
collection areas, approaching or exceeding the square-kilometre
scale under suitable observational conditions
\cite{LHAASO:2021ozi,Celli:2024cny}.
Unlike neutrinos, gamma rays do not suffer the same suppression
associated with a small interaction probability in the detector.
 A quantitative prediction nevertheless
requires convolution with the energy-dependent detector effective area and exposure.

Propagation further distinguishes the two messengers.
Very-high-energy gamma rays from the Galactic Center can be
attenuated through pair production on ambient radiation fields,
$\gamma\gamma\to e^+e^-$, with an energy-dependent survival
probability $\exp[-\tau_{\gamma\gamma}(E_\gamma,D)]$.
The absorbed energy may subsequently be redistributed through
electromagnetic cascades
\cite{Moskalenko:2005ng,Popescu:2017dgo}.
Neutrinos, by contrast, propagate essentially unattenuated
over Galactic distances. For nearby neutron stars, gamma-ray
attenuation is generally reduced. 



Finally, the nearby-source scenario should be distinguished
from the more robust prediction associated with the integrated
Galactic-center neutron star population. No neutron star is
currently known within the subparsec distance required for
a local source to match the benchmark Galactic-center capture
flux. Nevertheless, old and electromagnetically faint neutron
stars may remain unidentified. If dark matter accumulation
leads to repeated black hole formation and evaporation,
the resulting signal could consist of recurrent short bursts
from the same sky position. Depending on the repetition rate
and detector sensitivity, such emission may instead appear
effectively steady over long observational exposures.


\subsection{Comparison with celestial dark matter annihilation signals}
\label{sec:comparison-ann}

Having discussed the characteristic luminosity and observational
prospects, we now turn to the spectral and spatial morphology of the
signal, which provides additional information for its identification and
observational characterization. In particular, we ask whether these
observables can distinguish mediators produced through Hawking evaporation
from those generated by alternative mechanisms. As a representative
comparison, we consider monochromatic mediator production through dark
matter annihilation.

A central result of Sec.~\ref{sec:angular} is that, in the
ultra-relativistic regime and for
$R_{\rm NS}\ll\ell_{\rm dec}\ll D$, the normalized energy-integrated
angular profile is approximately independent of the mediator energy and,
consequently, of the black hole temperature. Although the Hawking
temperature increases during evaporation and continuously modifies the
mediator energy spectrum, the angular profile remains approximately
unchanged. This follows from the compensation between the boost-enhanced
decay length, $\ell_{\rm dec}\propto\gamma_S$, and relativistic beaming,
$\alpha_{\rm beam}\propto\gamma_S^{-1}$. The energy-resolved angular
distribution, however, retains information about the mediator energy and
the Hawking spectrum.

The energy and angular distributions therefore provide complementary
information about the production and propagation of the long-lived
mediator. To illustrate this, we contrast Hawking emission from a
microscopic black hole with monochromatic mediator production, as can
arise from nonrelativistic dark matter annihilation,
$\chi\chi\to SS$ \cite{Pospelov:2008jd,Batell:2009zp,Bell:2011sn,
Feng:2016ijc,Kouvaris:2016ltf,Leane:2017vag}. The two mechanisms produce markedly different mediator
spectra: black hole evaporation generates a broad spectrum that evolves
during evaporation, whereas dark matter annihilation produces an
approximately monochromatic spectrum,
\begin{equation}
 \frac{dN_S^{\rm ann}}{dE_S}
 \propto \delta(E_S-E_0).
 \label{eq:ann-spectrum}
\end{equation}

%
%

\begin{figure*}[!t]
    \centering
    \begin{subfigure}[t]{0.48\textwidth}
        \centering
        \includegraphics[width=\textwidth]{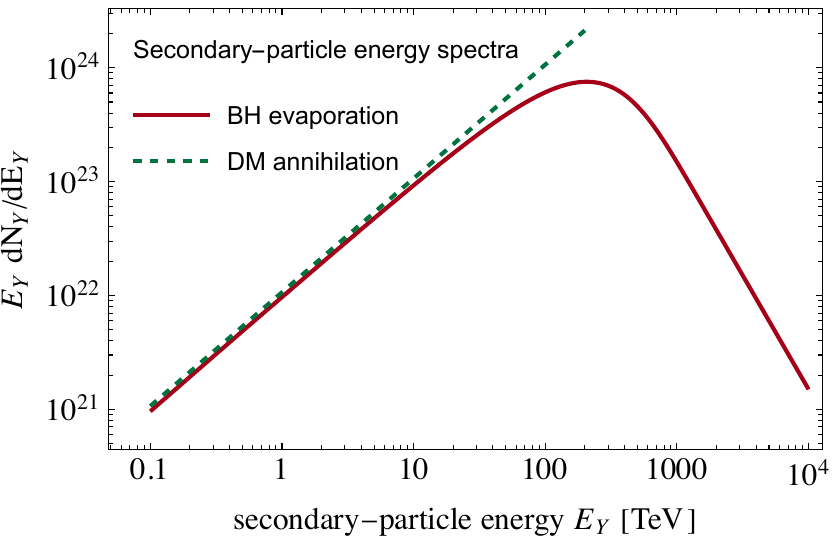}
    \end{subfigure}
    \hfill
    \begin{subfigure}[t]{0.48\textwidth}
        \centering
        \includegraphics[width=\textwidth]{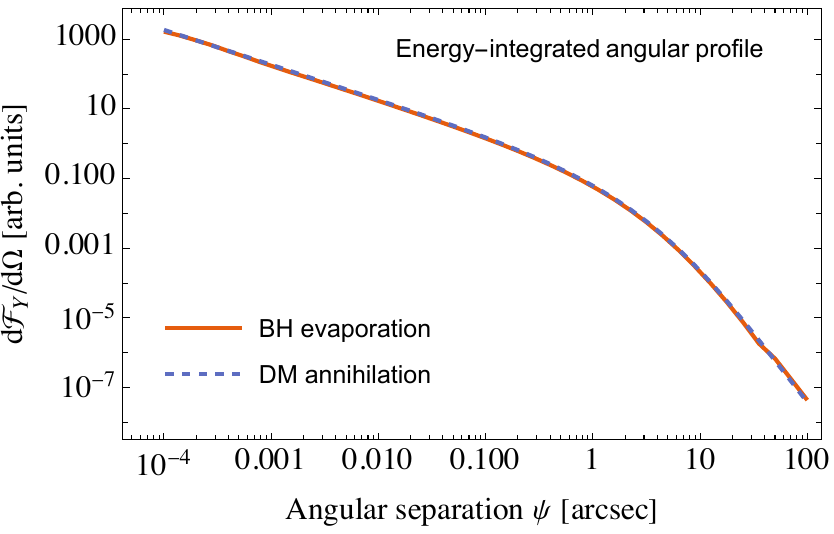}
    \end{subfigure}

\par\vspace{0.4cm}

    \begin{subfigure}[t]{0.48\textwidth}
        \centering
        \includegraphics[width=\textwidth]{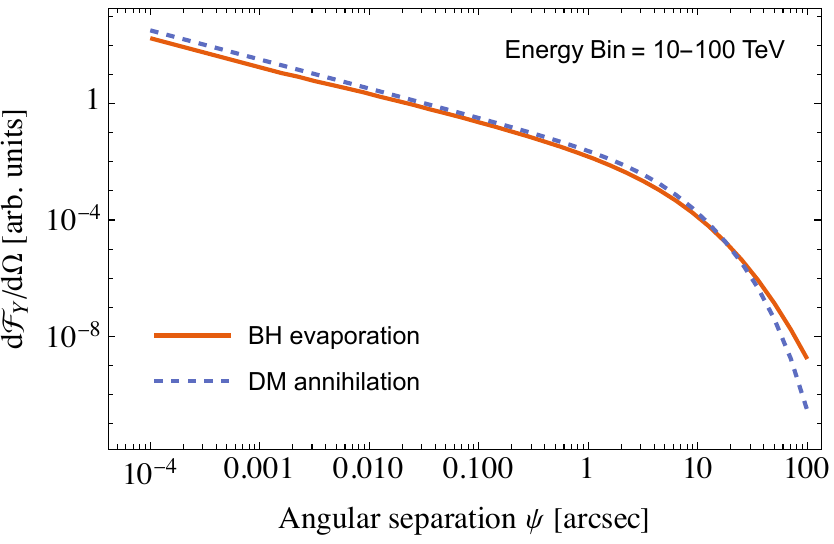}
    \end{subfigure}
    \hfill
    \begin{subfigure}[t]{0.48\textwidth}
        \centering
        \includegraphics[width=\textwidth]{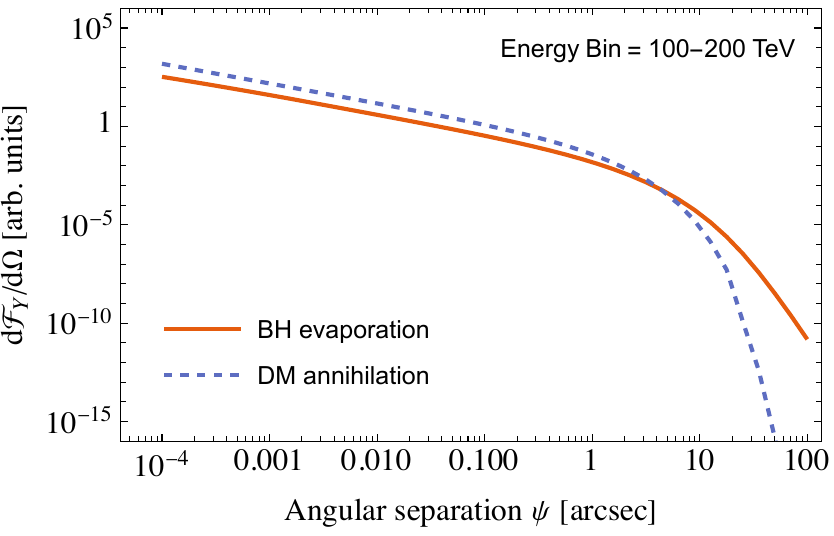}
    \end{subfigure}
    \caption{\small{Comparison of the secondary-particle spectra and angular distributions
from black hole evaporation and dark matter annihilation.
The upper-left panel shows the secondary-particle energy spectra, while
the upper-right panel shows the corresponding energy-integrated angular
profiles. The lower panels show the angular distributions
restricted to the secondary-energy intervals  10--100 TeV (lower left)
and  100--200 TeV (lower right).
The two scenarios are normalized to the same total number of emitted
mediators. The benchmark parameters are
$D=10~{\rm pc}$, $m_S=1~{\rm TeV}$, and $\tau_S=10^4~{\rm s}$, with
$T_{\rm BH}^{\rm init}=100~{\rm TeV}$ for black hole evaporation and a monochromatic
mediator energy $E_0=200~{\rm TeV}$ for dark matter annihilation.
The overall normalization of the angular distributions is omitted.}}
    \label{fig:BH-ann-comparison}
\end{figure*}

%
%

As an illustrative benchmark, shown in
Fig.~\ref{fig:BH-ann-comparison}, we consider a nearby neutron star and
adopt the following distance and mediator parameters  $D=10~{\rm pc}$,
 $m_S=1~{\rm TeV}$,
 $\tau_S=10^4~{\rm s}$,
together with an initial Hawking temperature
$T_{\rm BH}^{\rm init}=100~{\rm TeV}$. For the monochromatic source
scenario, we take $E_0=200~{\rm TeV}$. Such macroscopic lifetimes are
representative of the weakly coupled BSM scenarios discussed in
Sec.~\ref{sec:Escape}, although the precise relation between $m_S$,
$\tau_S$, and the underlying microscopic couplings is model dependent.
The benchmark should therefore be regarded as illustrative of the propagation and decay phenomenology, rather than as a specific particle model. It should not be regarded as a choice optimized for observational prospects either.

Figure~\ref{fig:BH-ann-comparison} compares the resulting secondary
energy and angular distributions for this benchmark. To facilitate a
comparison of their spectral and angular shapes, the two production
scenarios are normalized to the same total number of emitted mediators.
Their physical normalizations are, however, generally different and
depend, respectively, on the fraction $f_{S|H}$ of the Hawking
luminosity emitted into $S$ and on the branching fraction
${\rm Br}(\chi\chi\to SS)$ for dark matter annihilation.

The upper-left panel of Fig.~\ref{fig:BH-ann-comparison} compares the
secondary-particle energy spectra. In both cases, the secondary spectrum
is obtained by convolving the two-body decay kernel with the parent
mediator spectrum. For a monochromatic source, the mediator spectrum is given by
Eq.~\eqref{eq:ann-spectrum}, and the convolution reduces to the
box-like secondary spectrum 
\begin{equation}
\frac{dN_Y}{dE_Y}
=
\frac{2}{E_0\beta_0}\,
\Theta(E_Y-E_Y^-)\,
\Theta(E_Y^+-E_Y),
\qquad
E_Y^\pm=\frac{E_0}{2}(1\pm\beta_0),
\label{eq:ann-endpoints}
\end{equation}
where
$\beta_0=(1-m_S^2/E_0^2)^{1/2}$.
For black hole evaporation, the same decay kernel is instead convolved
with the continuous, time-integrated Hawking spectrum, producing a much
broader secondary spectrum given by Eq. \eqref{eq:secondary}. The spectral shape therefore provides a
direct discriminator between the two production mechanisms.

A different behavior emerges after integrating over the secondary energy.
As shown in the upper-right panel of Fig.~\ref{fig:BH-ann-comparison},
the normalized angular profiles become 
 indistinguishable despite
the very different mediator energy spectra. As derived in
Sec.~\ref{sec:angular}, the characteristic transverse displacement is
\begin{equation}
 \ell_{\rm dec}\alpha
 \sim
 \left(\gamma_S c\tau_S\right)\gamma_S^{-1}
 \sim c\tau_S,
\end{equation}
and is therefore approximately independent of the mediator boost. The
corresponding angular scale is
\begin{equation}
 \psi_{\rm char}\sim\frac{c\tau_S}{D},
 \label{eq:psi-char}
\end{equation}
provided $R_{\rm NS}\ll\ell_{\rm dec}\ll D$. For the benchmark of
Fig.~\ref{fig:BH-ann-comparison}, $c\tau_S/D\simeq2~{\rm arcsec}$.

More precisely, in terms of the dimensionless variable
$\xi=D\psi/(c\tau_S)$ introduced in
Eq.~\eqref{eq:angular-variables}, the normalized energy-integrated angular
profile approaches the universal form ${\cal A}(\xi)$ of
Eq.~\eqref{eq:Aangular}. Its dependence on $\gamma_S$ disappears when kinematic energy boundaries can be neglected. This explains the apparent coincidence
of the black hole and dark matter annihilation curves in
Fig.~\ref{fig:BH-ann-comparison}. Hence, the energy-integrated angular morphology
primarily probes $\tau_S/D$, rather than the mediator energy or production
spectrum. For a source at known distance, it therefore provides a direct
probe of the mediator lifetime.

The degeneracy is broken once the secondary energy is retained. The lower
panels of Fig.~\ref{fig:BH-ann-comparison} show the angular distributions
in the $100$--$200~{\rm TeV}$ and $10$--$100~{\rm TeV}$ energy intervals.
The black hole and monochromatic-source profiles now differ because
restricting $E_Y$ selects different regions of mediator energy and decay
kinematics. The effect is particularly pronounced near the high-energy
endpoint of the monochromatic spectrum.

\begin{table}[t]
\centering
\small
\setlength{\tabcolsep}{4pt}
\renewcommand{\arraystretch}{1.1}
\begin{tabular}{lccccc}
\hline\hline
Mediator &
$m_S$ &
$c\tau_S$ &
$\ell_{\rm dec}$ &
$\psi_{\rm NS}$ ($D=120~{\rm pc}$) &
$\psi_{\rm GC}$ ($D=8.3~{\rm kpc}$) \\
\hline
Gravitational $X_1$ &
$200~{\rm GeV}$ &
$0.24~{\rm pc}$ &
$1.2\times10^2~{\rm pc}$ &
--- &
$1.6\times10^{-3}\,{\rm deg}$ \\
Gravitational $X_2$ &
$10^3~{\rm GeV}$ &
$1.9\times10^{-3}~{\rm pc}$ &
$0.19~{\rm pc}$ &
$9.1\times10^{-4}\,{\rm deg}$ &
$1.3\times10^{-5}\,{\rm deg}$ \\
ALP $a_1$ &
$10^3~{\rm GeV}$ &
$9.7\times10^{-2}~{\rm pc}$ &
$9.7~{\rm pc}$ &
$4.6\times10^{-2}\,{\rm deg}$ &
$6.7\times10^{-4}\,{\rm deg}$ \\
ALP $a_2$ &
$20 ~{\rm TeV}$ &
$1.46\times10^2~{\rm pc}$ &
$7.1\times10^2~{\rm pc}$ &
--- &
$\sim1.0\,{\rm deg}$ \\
\hline
\end{tabular}
\caption{
Illustrative, phenomenologically motivated benchmarks
for long-lived mediators $S$, assuming
$T_{\rm BH}^{\rm init}=\mathcal{O}(100)~{\rm TeV}$. The lab-frame decay lengths $\ell_{\rm dec}$ are evaluated
at a reference mediator energy $E_S=100~{\rm TeV}$.
The characteristic angular scales are given for
a nearby neutron star at $D_{\rm NS}=120~{\rm pc}$
and the Galactic Center at $D_{\rm GC}=8.3~{\rm kpc}$.
For gravitationally coupled scalars, we assume
$\tilde c=1$.
The angular estimates use
$\psi_{\rm char}\simeq c\tau_S/D$
and are valid in the regime $\ell_{\rm dec}\ll D$.
A dash indicates that this approximation is not applicable.
}
\label{tab:mediator-angular-benchmarks}
\end{table}

\subsection{Observational Prospects}
\label{sec:observational-prospects}

The joint energy--angular distribution of
Eq.~\eqref{eq:fluence-def} contains information that is lost
upon integration over either variable. The energy spectrum
probes the production mechanism and characteristic mediator
energies, whereas the energy-integrated angular profile is
primarily sensitive to the mediator lifetime and propagation.
Energy-binned angular profiles retain both types of
information and may therefore help distinguish scenarios
with nearly degenerate integrated spectra or angular
morphologies.

The observability of the angular morphology depends on its
characteristic scale relative to the detector angular
resolution, as well as on the available event statistics.
For relativistic mediators, the characteristic angular
scale is approximately
$\psi_{\rm char}\sim c\tau_S/D$,
provided that
$R_{\rm NS}\ll\ell_{\rm dec}\ll D$.
Since $\ell_{\rm dec}\simeq\gamma_S c\tau_S$,
this condition also implies
$\psi_{\rm char}\ll1/\gamma_S$.
Consequently, a sizable angular extension requires
a sufficiently long proper lifetime together with a
moderate Lorentz boost. When the decay length becomes
comparable to the source distance, the finite-distance
geometry must be taken into account.

To illustrate the range of possible angular signatures,
Table~\ref{tab:mediator-angular-benchmarks} presents
representative benchmarks for two gravitationally coupled
scalars, $X_1$ and $X_2$, and two axion-like particles,
$a_1$ and $a_2$. We consider both a nearby neutron star
at $D_{\rm}=120~{\rm pc}$ and the Galactic Center
at $D_{\rm GC}=8.3~{\rm kpc}$, taking a representative
mediator energy $E_S=100~{\rm TeV}$.
The benchmarks span regimes ranging from compact,
effectively unresolved emission to potentially
extended photon halos. The angular scales listed in the table represent characteristic estimates rather than precise measures of the angular extent of the emission.

For gravitationally coupled scalar mediators, the
decay width is determined by the mediator mass and
the scale suppressing its interactions. At fixed
energy, the decay length scales approximately as
$\ell_{\rm dec}\propto m_X^{-4}$.
For the heavier benchmark, $m_{X_2}=1~{\rm TeV}$,
the decay length is much smaller than either
source distance, resulting in a very compact
angular distribution. For the lighter benchmark,
$m_{X_1}=200~{\rm GeV}$, the decay length becomes
comparable to the distance of the nearby neutron
star. In this case, the approximate scaling
$\psi_{\rm char}\sim c\tau_X/D_{\rm NS}$
is no longer applicable, and the finite radial
integration limits in
Eq.~\eqref{eq:angular-spectrum}
must be retained. At the Galactic Center, however,
both gravitational scalar benchmarks produce
angular extensions much smaller than the
resolution of current gamma-ray observatories.

Axion-like mediators offer greater flexibility, since
their masses and decay widths can be varied
independently through the relevant couplings.
In particular, the heavier ALP benchmark
$a_2$, with $m_{a_2}=20~{\rm TeV}$ and
$\tau_{a_2}=1.5\times10^{10}~{\rm s}$,
has a characteristic angular scale of order
$1^\circ$ at the Galactic Center.
This is larger than the angular radius
$\theta_{\rm NS}\simeq0.7^\circ$
corresponding to a neutron star population
confined within approximately $100~{\rm pc}$
of the Galactic Center. Such a mediator could
therefore generate a photon halo extending beyond
the underlying source population. 

If several long-lived mediator species carry
Hawking-emitted energy out of the host star,
their different lifetimes and decay kinematics
may generate distinct angular components.
A sufficiently resolved multi-component
morphology could therefore provide additional
information about the particle content of the
emitted radiation. In practice, identifying
such components would require adequate photon
statistics and angular resolution.

Nearby neutron stars provide a complementary
setting for investigating mediator propagation.
Their smaller distances may enhance the apparent
angular extension and reduce gamma-ray attenuation.
However, the substantially lower dark matter
density in the Solar neighborhood also reduces
the capture power and hence the expected signal.
Unless exceptionally close, individual nearby
neutron stars may therefore be less favorable
for detection than the integrated Galactic Center
population. 
Finally, the motion of the host neutron star may
introduce additional observational effects, such
as Doppler shifts and aberration, which are not
included in our analysis.

Ultimately, a multi-messenger approach
combining gamma-ray and neutrino observations with
infrared monitoring of neutron stars
\cite{Raj:2024kjq,Saha:2025fgu}
could improve the prospects for detecting the faint
signatures of long-lived particles carrying
Hawking-emitted energy away from their host stars.
Even when the intrinsic angular extension is too
small to be resolved, a sufficiently strong signal
could appear as a point-like source and potentially
be localized and associated with a known neutron star.

\section{Conclusions}
\label{sec:conclusions}

Microscopic black holes potentially formed inside
neutron stars could provide an astrophysical setting
for the realization of Hawking evaporation.
This fundamental quantum-gravitational phenomenon,
which remains observationally unconfirmed,
could become accessible through long-lived particles
emitted by the black holes that escape the dense
stellar medium and subsequently decay outside the star.
Their decays may give rise to high-energy neutrino,
photon, and charged-particle signals, rendering
otherwise hidden black hole evaporation observable
through a faint but potentially detectable secondary flux.
 A characteristic feature
of this signal is its broad energy spectrum, extending to
ultra-high energies as a consequence of the high Hawking
temperatures of microscopic black holes. This motivates the study
of its high-energy flux~\cite{Dalianis:2026jcl} and, in the
present work, its angular properties.

In this work, we have examined the secondary energy spectrum and derived
the joint energy--angular distribution, Eq.~(\ref{eq:master}), allowing the
spectral and spatial properties of the signal to be studied simultaneously.
A central result is that, in the relativistic regime and for
$R_{\rm NS}\ll\ell_{\rm dec}\ll D$, the normalized energy-integrated
angular profile becomes approximately independent of the mediator energy
and, consequently, of the black hole temperature. This behavior results
from the compensation between relativistic beaming, which narrows the
decay cone as $\alpha\sim\gamma_S^{-1}$, and the increasing decay length,
$\ell_{\rm dec}\simeq\gamma_S c\tau_S$. 
The angular profile therefore remains approximately unchanged
as $T_{\rm BH}$ increases during evaporation, with its
characteristic scale controlled primarily by the mediator
proper lifetime and source distance. 
For a source at known distance, a resolved angular extension
could thus constrain the mediator lifetime.

The approximate independence of the angular profile from the mediator
energy also has implications for distinguishing different production
mechanisms. This is illustrated in Fig.~\ref{fig:BH-ann-comparison},
where black-hole evaporation is compared with monochromatic mediator
production from dark matter annihilation within celestial objects. The latter provides an alternative mechanism
for generating very-high-energy signals, provided that the dark matter mass
is sufficiently large, making it important to identify observables that can
distinguish between the two origins.
Although the two mechanisms produce markedly different
secondary-energy spectra, their normalized
energy-integrated angular profiles are nearly identical
in the relativistic regime, reflecting the weak dependence
of the angular morphology on the mediator energy.
This degeneracy can be lifted by retaining the secondary
energy: the energy-binned angular profiles differ,
particularly near the kinematic endpoint of the
monochromatic spectrum.
 Moreover, if microscopic
black holes form repeatedly, the recurrence of short evaporation bursts
from the same neutron star could provide a further temporal discriminator,
complementary to the spectral and angular signatures.

The generality and feasibility of this escape mechanism are
illustrated through several representative long-lived-particle
realizations, including gravitationally coupled scalars,
axion-like particles, dark photons, dark-$Z$ bosons, and
heavy neutral leptons. Their different masses, couplings,
decay modes, and lifetimes give rise to a broad range of
secondary spectra and propagation scales. Depending on
these properties, the escaping Hawking radiation may
give rise to neutrino, photon, charged-particle, or
multi-messenger signals, appearing as either unresolved
or spatially extended sources. In particular, sufficiently
long-lived mediators may generate photon halos extending
beyond the underlying neutron star population, as
illustrated by our Galactic Center ALP benchmark.
The spectral shape probes the production and decay
mechanism, while a resolved angular profile provides
complementary information on the mediator lifetime.
Furthermore, if several long-lived species contribute
to the Hawking radiation, their different lifetimes
may generate multiple characteristic angular scales,
potentially revealing the presence of more than one
mediator species. Finally, the framework developed
here can be generalized beyond neutron stars to
other stellar environments in which microscopic black
holes may form, such as white dwarfs or main-sequence
stars.

The absolute signal normalization remains subject to
both particle-physics and astrophysical uncertainties.
In particular, the population-level flux depends on
the dark matter distribution, neutron star population,
and capture efficiency, with substantial uncertainties
toward the Galactic Center. Additional model dependence
arises from the fraction of Hawking luminosity carried
by long-lived states and from the conditions governing
dark matter collapse and black hole formation.
 Ultimately, the predicted flux alone does not determine
the observability of the signal, and a quantitative assessment of the
detection prospects requires folding the predicted spectra with the
appropriate detector response. 
Among the possible secondary messengers, gamma rays
may offer favorable detection prospects owing to
the large effective areas and angular resolution of
gamma-ray observatories. 
Neutrinos, on the other hand, propagate essentially
unattenuated over Galactic distances.
A more detailed analysis of the expected neutrino
energy fluxes, event rates, and detection prospects
is presented in Ref.~\cite{Dalianis:2026jcl}.
Given the generally faint expected signals and their often
small angular extent, resolving the spatial morphology
may require exceptionally nearby sources or future
improvements in instrumental sensitivity and angular
resolution.  
Nevertheless, even an unresolved signal could
potentially be localized and associated with
a known neutron star, provided that sufficient
statistics are available. 

In summary, the energy--angular structure derived
here 
provides a characteristic signature of long-lived
particles emitted through microscopic black hole evaporation, and is independent of the absolute signal normalization.
Its  form encodes
information about both the properties of the particles
escaping the neutron star and their subsequent decays
into observable secondaries.
Overall, the prospect of probing the quantum properties
of black holes provides strong motivation for exploring
these hypothetical signatures, despite the considerable
observational challenges.
Multi-messenger searches targeting nearby neutron stars
and neutron-star-rich regions may therefore open a
new, albeit dim, window onto Hawking evaporation,
while simultaneously probing the properties of the
beyond the Standard Model particles that carry Hawking-emitted energy out
of their host stars.

%
%
%
%
\vspace{0.5cm}
\bibliographystyle{jhep}
\bibliography{references}

@article{Kouvaris:2016ltf,
    author = "Kouvaris, Chris and Lang{\ae}ble, Kasper and Nielsen, Niklas Gr{\o}nlund",
    title = "{The Spectrum of Darkonium in the Sun}",
    eprint = "1607.00374",
    archivePrefix = "arXiv",
    primaryClass = "hep-ph",
    reportNumber = "CP3-ORIGINS-2016-029",
    doi = "10.1088/1475-7516/2016/10/012",
    journal = "JCAP",
    volume = "10",
    pages = "012",
    year = "2016"
}

@article{IceCube:2013low,
    author = "Aartsen, M. G. and others",
    collaboration = "IceCube",
    title = "{Evidence for High-Energy Extraterrestrial Neutrinos at the IceCube Detector}",
    eprint = "1311.5238",
    archivePrefix = "arXiv",
    primaryClass = "astro-ph.HE",
    doi = "10.1126/science.1242856",
    journal = "Science",
    volume = "342",
    pages = "1242856",
    year = "2013"
}

@article{IceCube:2014stg,
    author = "Aartsen, M. G. and others",
    collaboration = "IceCube",
    title = "{Observation of High-Energy Astrophysical Neutrinos in Three Years of IceCube Data}",
    eprint = "1405.5303",
    archivePrefix = "arXiv",
    primaryClass = "astro-ph.HE",
    doi = "10.1103/PhysRevLett.113.101101",
    journal = "Phys. Rev. Lett.",
    volume = "113",
    pages = "101101",
    year = "2014"
}

@article{Hawking:1975vcx,
    author = "Hawking, S. W.",
    title = "{Particle Creation by Black Holes}",
    doi = "10.1007/BF02345020",
    journal = "Commun. Math. Phys.",
    volume = "43",
    pages = "199--220",
    year = "1975",
    note = "[Erratum: Commun. Math. Phys. \textbf{46}, 206 (1976)]"
}

@article{Hawking:1974rv,
    author = "Hawking, S. W.",
    title = "{Black hole explosions}",
    doi = "10.1038/248030a0",
    journal = "Nature",
    volume = "248",
    pages = "30--31",
    year = "1974"
}

@article{Dimopoulos:1982cz,
    author = "Dimopoulos, Savas and Preskill, John and Wilczek, Frank",
    title = "{Catalyzed Nucleon Decay in Neutron Stars}",
    reportNumber = "HUTP-82-A047, NSF-ITP-82-91",
    doi = "10.1016/0370-2693(82)90679-7",
    journal = "Phys. Lett. B",
    volume = "119",
    pages = "320",
    year = "1982"
}

@article{MacGibbon:1990zk,
    author = "MacGibbon, J. H. and Webber, B. R.",
    title = "{Quark and gluon jet emission from primordial black holes: The instantaneous spectra}",
    doi = "10.1103/PhysRevD.41.3052",
    journal = "Phys. Rev. D",
    volume = "41",
    pages = "3052--3079",
    year = "1990"
}

@article{Nakamura:2006uc,
    author = "Nakamura, Shuntaro and Yamaguchi, Masahiro",
    title = "{Gravitino production from heavy moduli decay and cosmological moduli problem revived}",
    eprint = "hep-ph/0602081",
    archivePrefix = "arXiv",
    reportNumber = "TU-765",
    doi = "10.1016/j.physletb.2006.05.078",
    journal = "Phys. Lett. B",
    volume = "638",
    pages = "389--395",
    year = "2006"
}

@article{Endo:2006zj,
    author = "Endo, Motoi and Hamaguchi, Koichi and Takahashi, Fuminobu",
    title = "{Moduli-induced gravitino problem}",
    eprint = "hep-ph/0602061",
    archivePrefix = "arXiv",
    reportNumber = "DESY-06-014",
    doi = "10.1103/PhysRevLett.96.211301",
    journal = "Phys. Rev. Lett.",
    volume = "96",
    pages = "211301",
    year = "2006"
}

@article{Cicoli:2015ylx,
    author = "Cicoli, Michele and Quevedo, Fernando and Valandro, Roberto",
    title = "{De Sitter from T-branes}",
    eprint = "1512.04558",
    archivePrefix = "arXiv",
    primaryClass = "hep-th",
    doi = "10.1007/JHEP03(2016)141",
    journal = "JHEP",
    volume = "03",
    pages = "141",
    year = "2016"
}

@article{Davoudiasl:2012ag,
    author = "Davoudiasl, Hooman and Lee, Hye-Sung and Marciano, William J.",
    title = "{'Dark' Z implications for Parity Violation, Rare Meson Decays, and Higgs Physics}",
    eprint = "1203.2947",
    archivePrefix = "arXiv",
    primaryClass = "hep-ph",
    doi = "10.1103/PhysRevD.85.115019",
    journal = "Phys. Rev. D",
    volume = "85",
    pages = "115019",
    year = "2012"
}

@article{Davoudiasl:2014kua,
    author = "Davoudiasl, Hooman and Lee, Hye-Sung and Marciano, William J.",
    title = "{Muon $g-2$ rare kaon decays, and parity violation from dark bosons}",
    eprint = "1402.3620",
    archivePrefix = "arXiv",
    primaryClass = "hep-ph",
    reportNumber = "JLAB-THY-14-1841",
    doi = "10.1103/PhysRevD.89.095006",
    journal = "Phys. Rev. D",
    volume = "89",
    number = "9",
    pages = "095006",
    year = "2014"
}

@article{McDermott:2011jp,
    author = "McDermott, Samuel D. and Yu, Hai-Bo and Zurek, Kathryn M.",
    title = "{Constraints on Scalar Asymmetric Dark Matter from Black Hole Formation in Neutron Stars}",
    eprint = "1103.5472",
    archivePrefix = "arXiv",
    primaryClass = "hep-ph",
    reportNumber = "MCTP-11-16",
    doi = "10.1103/PhysRevD.85.023519",
    journal = "Phys. Rev. D",
    volume = "85",
    pages = "023519",
    year = "2012"
}

@article{Nguyen:2022zwb,
    author = "Nguyen, Thong T. Q. and Tait, Tim M. P.",
    title = "{Bounds on long-lived dark matter mediators from neutron stars}",
    eprint = "2212.12547",
    archivePrefix = "arXiv",
    primaryClass = "hep-ph",
    doi = "10.1103/PhysRevD.107.115016",
    journal = "Phys. Rev. D",
    volume = "107",
    number = "11",
    pages = "115016",
    year = "2023"
}

@article{Goldman:1989nd,
    author = "Goldman, I. and Nussinov, S.",
    title = "{Weakly Interacting Massive Particles and Neutron Stars}",
    doi = "10.1103/PhysRevD.40.3221",
    journal = "Phys. Rev. D",
    volume = "40",
    pages = "3221--3230",
    year = "1989"
}

@article{Page:1976wx,
    author = "Page, Don N. and Hawking, S. W.",
    title = "{Gamma rays from primordial black holes}",
    doi = "10.1086/154350",
    journal = "Astrophys. J.",
    volume = "206",
    pages = "1--7",
    year = "1976"
}

@article{Navarro:1995iw,
    author = "Navarro, Julio F. and Frenk, Carlos S. and White, Simon D. M.",
    title = "{The Structure of cold dark matter halos}",
    eprint = "astro-ph/9508025",
    archivePrefix = "arXiv",
    doi = "10.1086/177173",
    journal = "Astrophys. J.",
    volume = "462",
    pages = "563--575",
    year = "1996"
}

@article{Navarro:1996gj,
    author = "Navarro, Julio F. and Frenk, Carlos S. and White, Simon D. M.",
    title = "{A Universal density profile from hierarchical clustering}",
    eprint = "astro-ph/9611107",
    archivePrefix = "arXiv",
    doi = "10.1086/304888",
    journal = "Astrophys. J.",
    volume = "490",
    pages = "493--508",
    year = "1997"
}

@article{Bell:2013xk,
    author = "Bell, Nicole F. and Melatos, Andrew and Petraki, Kalliopi",
    title = "{Realistic neutron star constraints on bosonic asymmetric dark matter}",
    eprint = "1301.6811",
    archivePrefix = "arXiv",
    primaryClass = "hep-ph",
    reportNumber = "NIKHEF-2013-002",
    doi = "10.1103/PhysRevD.87.123507",
    journal = "Phys. Rev. D",
    volume = "87",
    number = "12",
    pages = "123507",
    year = "2013"
}

@article{Autzen:2014tza,
    author = "Autzen, Martin and Kouvaris, Chris",
    title = "{Blocking the Hawking Radiation}",
    eprint = "1403.1072",
    archivePrefix = "arXiv",
    primaryClass = "astro-ph.SR",
    doi = "10.1103/PhysRevD.89.123519",
    journal = "Phys. Rev. D",
    volume = "89",
    number = "12",
    pages = "123519",
    year = "2014"
}

@article{Kouvaris:2013kra,
    author = "Kouvaris, Chris and Tinyakov, Peter",
    title = "{Growth of Black Holes in the interior of Rotating Neutron Stars}",
    eprint = "1312.3764",
    archivePrefix = "arXiv",
    primaryClass = "astro-ph.SR",
    doi = "10.1103/PhysRevD.90.043512",
    journal = "Phys. Rev. D",
    volume = "90",
    number = "4",
    pages = "043512",
    year = "2014"
}

@article{Giffin:2021kgb,
    author = "Giffin, Pierce and Lloyd, John and McDermott, Samuel D. and Profumo, Stefano",
    title = "{Neutron star quantum death by small black holes}",
    eprint = "2105.06504",
    archivePrefix = "arXiv",
    primaryClass = "hep-ph",
    reportNumber = "FERMILAB-PUB-21-259-T",
    doi = "10.1103/PhysRevD.105.123030",
    journal = "Phys. Rev. D",
    volume = "105",
    number = "12",
    pages = "123030",
    year = "2022"
}

@article{Kouvaris:2011fi,
    author = "Kouvaris, Chris and Tinyakov, Peter",
    title = "{Excluding Light Asymmetric Bosonic Dark Matter}",
    eprint = "1104.0382",
    archivePrefix = "arXiv",
    primaryClass = "astro-ph.CO",
    doi = "10.1103/PhysRevLett.107.091301",
    journal = "Phys. Rev. Lett.",
    volume = "107",
    pages = "091301",
    year = "2011"
}

@article{Saha:2025fgu,
    author = "Saha, Akash Kumar and Dubey, Abhishek and Raj, Nirmal",
    title = "{Hawking heating of neutron stars by dark matter}",
    eprint = "2511.19599",
    archivePrefix = "arXiv",
    primaryClass = "hep-ph",
    doi = " ",
    journal = "",
    volume = "",
    pages = "",
    year = "2025"
}

@article{KM3Net:2016zxf,
    author = "Adrian-Martinez, S. and others",
    collaboration = "KM3Net",
    title = "{Letter of intent for KM3NeT 2.0}",
    eprint = "1601.07459",
    archivePrefix = "arXiv",
    primaryClass = "astro-ph.IM",
    doi = "10.1088/0954-3899/43/8/084001",
    journal = "J. Phys. G",
    volume = "43",
    number = "8",
    pages = "084001",
    year = "2016"
}

@article{Klipfel:2025jql,
    author = "Klipfel, Alexandra P. and Kaiser, David I.",
    title = "{Ultrahigh-Energy Neutrinos from Primordial Black Holes}",
    eprint = "2503.19227",
    archivePrefix = "arXiv",
    primaryClass = "hep-ph",
    reportNumber = "Preprint MIT-CTP/5852",
    doi = "10.1103/vnm4-7wdc",
    journal = "Phys. Rev. Lett.",
    volume = "135",
    number = "12",
    pages = "121003",
    year = "2025"
}

@article{Bramante:2023djs,
    author = "Bramante, Joseph and Raj, Nirmal",
    title = "{Dark matter in compact stars}",
    eprint = "2307.14435",
    archivePrefix = "arXiv",
    primaryClass = "hep-ph",
    doi = "10.1016/j.physrep.2023.12.001",
    journal = "Phys. Rept.",
    volume = "1052",
    pages = "1--48",
    year = "2024"
}

@article{Gould:1989gw,
    author = "Gould, Andrew and Draine, Bruce T. and Romani, Roger W. and Nussinov, Shmuel",
    title = "{Neutron Stars: Graveyard of Charged Dark Matter}",
    reportNumber = "IASSNS-AST 89/55",
    doi = "10.1016/0370-2693(90)91745-W",
    journal = "Phys. Lett. B",
    volume = "238",
    pages = "337--343",
    year = "1990"
}

@article{Carr:2020gox,
    author = "Carr, Bernard and Kohri, Kazunori and Sendouda, Yuuiti and Yokoyama, Jun'ichi",
    title = "{Constraints on primordial black holes}",
    eprint = "2002.12778",
    archivePrefix = "arXiv",
    primaryClass = "astro-ph.CO",
    reportNumber = "RESCEU-03/20; KEK-Cosmo-249; KEK-TH-2199; IPMU20-0024",
    doi = "10.1088/1361-6633/ac1e31",
    journal = "Rept. Prog. Phys.",
    volume = "84",
    number = "11",
    pages = "116902",
    year = "2021"
}

@article{Kouvaris:2010jy,
    author = "Kouvaris, Chris and Tinyakov, Peter",
    title = "{Constraining Asymmetric Dark Matter through observations of compact stars}",
    eprint = "1012.2039",
    archivePrefix = "arXiv",
    primaryClass = "astro-ph.HE",
    doi = "10.1103/PhysRevD.83.083512",
    journal = "Phys. Rev. D",
    volume = "83",
    pages = "083512",
    year = "2011"
}

@article{Kouvaris:2011gb,
    author = "Kouvaris, Chris",
    title = "{Limits on Self-Interacting Dark Matter}",
    eprint = "1111.4364",
    archivePrefix = "arXiv",
    primaryClass = "astro-ph.CO",
    doi = "10.1103/PhysRevLett.108.191301",
    journal = "Phys. Rev. Lett.",
    volume = "108",
    pages = "191301",
    year = "2012"
}

@article{Bramante:2013hn,
    author = "Bramante, Joseph and Fukushima, Keita and Kumar, Jason",
    title = "{Constraints on bosonic dark matter from observation of old neutron stars}",
    eprint = "1301.0036",
    archivePrefix = "arXiv",
    primaryClass = "hep-ph",
    doi = "10.1103/PhysRevD.87.055012",
    journal = "Phys. Rev. D",
    volume = "87",
    number = "5",
    pages = "055012",
    year = "2013"
}

@article{Bramante:2013nma,
    author = "Bramante, Joseph and Fukushima, Keita and Kumar, Jason and Stopnitzky, Elan",
    title = "{Bounds on self-interacting fermion dark matter from observations of old neutron stars}",
    eprint = "1310.3509",
    archivePrefix = "arXiv",
    primaryClass = "hep-ph",
    doi = "10.1103/PhysRevD.89.015010",
    journal = "Phys. Rev. D",
    volume = "89",
    number = "1",
    pages = "015010",
    year = "2014"
}

@article{Kolb:1982si,
    author = "Kolb, Edward W. and Colgate, Stirling A. and Harvey, Jeffrey A.",
    title = "{Monopole Catalysis of Nucleon Decay in Neutron Stars}",
    reportNumber = "LA-UR-82-1963",
    doi = "10.1103/PhysRevLett.49.1373",
    journal = "Phys. Rev. Lett.",
    volume = "49",
    pages = "1373",
    year = "1982"
}

@article{Carr:1976zz,
  author = {B. J. Carr},
  title = {Some cosmological consequences of primordial black-hole evaporations},
  journal = {Astrophys. J.},
  volume = {206},
  pages = {8},
  year = {1976}
}

@article{Dave:2019epr,
    author = "Dave, Pranav and Taboada, Ignacio",
    collaboration = "IceCube",
    title = "{Neutrinos from Primordial Black Hole Evaporation}",
    eprint = "1908.05403",
    archivePrefix = "arXiv",
    primaryClass = "astro-ph.HE",
    reportNumber = "PoS-ICRC2019-863",
    doi = "10.22323/1.358.0863",
    journal = "PoS",
    volume = "ICRC2019",
    pages = "863",
    year = "2021"
}

@article{Anchordoqui:2025xug,
    author = "Anchordoqui, Luis A. and Halzen, Francis and Lust, Dieter",
    title = "{Neutrinos from primordial black holes in theories with extra dimensions}",
    eprint = "2505.23414",
    archivePrefix = "arXiv",
    primaryClass = "hep-ph",
    reportNumber = "MPP-2025-110; LMU-ASC 14/25",
    doi = "10.1103/5kt2-5pvj",
    journal = "Phys. Rev. D",
    volume = "112",
    number = "8",
    pages = "083034",
    year = "2025"
}

@article{Acevedo:2020gro,
    author = "Acevedo, Javier F. and Bramante, Joseph and Goodman, Alan and Kopp, Joachim and Opferkuch, Toby",
    title = "{Dark Matter, Destroyer of Worlds: Neutrino, Thermal, and Existential Signatures from Black Holes in the Sun and Earth}",
    eprint = "2012.09176",
    archivePrefix = "arXiv",
    primaryClass = "hep-ph",
    reportNumber = "CERN-TH-2020-209, MITP/20-081",
    doi = "10.1088/1475-7516/2021/04/026",
    journal = "JCAP",
    volume = "04",
    pages = "026",
    year = "2021"
}

@article{Raj:2024kjq,
    author = "Raj, Nirmal and Shivanna, Prajwal and Rachh, Gaurav Niraj",
    title = "{Exploring reheated sub-40000 Kelvin neutron stars with JWST, ELT, and TMT}",
    eprint = "2403.07496",
    archivePrefix = "arXiv",
    primaryClass = "astro-ph.HE",
    doi = "10.1103/PhysRevD.109.123040",
    journal = "Phys. Rev. D",
    volume = "109",
    number = "12",
    pages = "123040",
    year = "2024"
}

@article{Acevedo:2024ttq,
    author = "Acevedo, Javier F. and Bramante, Joseph and Liu, Qinrui and Tyagi, Narayani",
    title = "{Neutrino and gamma-ray signatures of inelastic dark matter annihilating outside neutron stars}",
    eprint = "2404.10039",
    archivePrefix = "arXiv",
    primaryClass = "hep-ph",
    reportNumber = "SLAC-PUB-17767",
    doi = "10.1088/1475-7516/2025/03/028",
    journal = "JCAP",
    volume = "03",
    pages = "028",
    year = "2025"
}

@article{Ibarra:2012dw,
    author = "Ibarra, Alejandro and Lopez Gehler, Sergio and Pato, Miguel",
    title = "{Dark matter constraints from box-shaped gamma-ray features}",
    eprint = "1205.0007",
    archivePrefix = "arXiv",
    primaryClass = "hep-ph",
    doi = "10.1088/1475-7516/2012/07/043",
    journal = "JCAP",
    volume = "07",
    pages = "043",
    year = "2012"
}

@article{Leane:2021ihh,
    author = "Leane, Rebecca K. and Linden, Tim and Mukhopadhyay, Payel and Toro, Natalia",
    title = "{Celestial-Body Focused Dark Matter Annihilation Throughout the Galaxy}",
    eprint = "2101.12213",
    archivePrefix = "arXiv",
    primaryClass = "astro-ph.HE",
    doi = "10.1103/PhysRevD.103.075030",
    journal = "Phys. Rev. D",
    volume = "103",
    number = "7",
    pages = "075030",
    year = "2021"
}

@article{KM3NeT:2025npi,
    author = "Aiello, S. and others",
    collaboration = "KM3NeT",
    title = "{Observation of an ultra-high-energy cosmic neutrino with KM3NeT}",
    doi = "10.1038/s41586-024-08543-1",
    journal = "Nature",
    volume = "638",
    number = "8050",
    pages = "376--382",
    year = "2025",
    note = "[Erratum: Nature 640, E3 (2025)]"
}

@article{Bose:2021yhz,
    author = "Bose, Debajit and Maity, Tarak Nath and Ray, Tirtha Sankar",
    title = "{Neutrinos from captured dark matter annihilation in a galactic population of neutron stars}",
    eprint = "2108.12420",
    archivePrefix = "arXiv",
    primaryClass = "hep-ph",
    doi = "10.1088/1475-7516/2022/05/001",
    journal = "JCAP",
    volume = "05",
    number = "05",
    pages = "001",
    year = "2022"
}

@article{Bose:2024wsh,
    author = "Bose, Debajit and Pramanick, Rohan and Ray, Tirtha Sankar",
    title = "{Neutrinos from captured dark matter in galactic stars}",
    eprint = "2405.07894",
    archivePrefix = "arXiv",
    primaryClass = "hep-ph",
    doi = "10.1016/j.physletb.2025.139542",
    journal = "Phys. Lett. B",
    volume = "866",
    pages = "139542",
    year = "2025"
}

@article{Dalianis:2026jcl,
    author = "Dalianis, Ioannis",
    title = "{High-Energy Neutrinos from Black Hole Evaporation in Neutron Stars}",
    eprint = "2607.13755",
    archivePrefix = "arXiv",
    primaryClass = "hep-ph",
    month = "7",
    year = "2026"
}

@article{HESS:2006fka,
    author = "Aharonian, F. and others",
    collaboration = "H.E.S.S.",
    title = "{Observations of the Crab Nebula with H.E.S.S}",
    eprint = "astro-ph/0607333",
    archivePrefix = "arXiv",
    doi = "10.1051/0004-6361:20065351",
    journal = "Astron. Astrophys.",
    volume = "457",
    pages = "899--915",
    year = "2006"
}

@article{LHAASO:2021ozi,
    author = "Aharonian, F. and others",
    collaboration = "LHAASO",
    title = "{Performance of LHAASO-WCDA and observation of the Crab Nebula as a standard candle}",
    doi = "10.1088/1674-1137/ac041b",
    journal = "Chin. Phys. C",
    volume = "45",
    number = "8",
    pages = "085002",
    year = "2021"
}

@book{CTAConsortium:2017dvg,
    author = "Acharya, B. S. and others",
    collaboration = "CTA Consortium",
    title = "{Science with the Cherenkov Telescope Array}",
    eprint = "1709.07997",
    archivePrefix = "arXiv",
    primaryClass = "astro-ph.IM",
    doi = "10.1142/10986",
    isbn = "978-981-327-008-4",
    publisher = "WSP",
    month = "11",
    year = "2018"
}

@article{Ukwatta:2015iba,
    author = "Ukwatta, T. N. and Stump, D. R. and Linnemann, J. T. and MacGibbon, J. H. and Marinelli, S. S. and Yapici, T. and Tollefson, K.",
    title = "{Primordial Black Holes: Observational Characteristics of The Final Evaporation}",
    eprint = "1510.04372",
    archivePrefix = "arXiv",
    primaryClass = "astro-ph.HE",
    doi = "10.1016/j.astropartphys.2016.03.007",
    journal = "Astropart. Phys.",
    volume = "80",
    pages = "90--114",
    year = "2016"
}

@book{Misner:1973prb,
    author = "Misner, Charles W. and Thorne, K. S. and Wheeler, J. A.",
    title = "{Gravitation}",
    isbn = "978-0-7167-0344-0, 978-0-691-17779-3",
    publisher = "W. H. Freeman",
    address = "San Francisco",
    year = "1973"
}

@article{Hook:2021ous,
    author = "Hook, Anson and Marques-Tavares, Gustavo and Ristow, Clayton",
    title = "{Supernova constraints on an axion-photon-dark photon interaction}",
    eprint = "2105.06476",
    archivePrefix = "arXiv",
    primaryClass = "hep-ph",
    doi = "10.1007/JHEP06(2021)167",
    journal = "JHEP",
    volume = "06",
    pages = "167",
    year = "2021"
}

@article{Ilten:2018crw,
    author = "Ilten, Philip and Soreq, Yotam and Williams, Mike and Xue, Wei",
    title = "{Serendipity in dark photon searches}",
    eprint = "1801.04847",
    archivePrefix = "arXiv",
    primaryClass = "hep-ph",
    reportNumber = "MIT-CTP/4976, CERN-TH-2017-282, MIT-CTP-4976",
    doi = "10.1007/JHEP06(2018)004",
    journal = "JHEP",
    volume = "06",
    pages = "004",
    year = "2018"
}

@article{Pospelov:2008jd,
    author = "Pospelov, Maxim and Ritz, Adam",
    title = "{Astrophysical Signatures of Secluded Dark Matter}",
    eprint = "0810.1502",
    archivePrefix = "arXiv",
    primaryClass = "hep-ph",
    doi = "10.1016/j.physletb.2008.12.012",
    journal = "Phys. Lett. B",
    volume = "671",
    pages = "391--397",
    year = "2009"
}

@article{Batell:2009zp,
    author = "Batell, Brian and Pospelov, Maxim and Ritz, Adam and Shang, Yanwen",
    title = "{Solar Gamma Rays Powered by Secluded Dark Matter}",
    eprint = "0910.1567",
    archivePrefix = "arXiv",
    primaryClass = "hep-ph",
    doi = "10.1103/PhysRevD.81.075004",
    journal = "Phys. Rev. D",
    volume = "81",
    pages = "075004",
    year = "2010"
}

@article{Bell:2011sn,
    author = "Bell, Nicole F. and Petraki, Kalliopi",
    title = "{Enhanced neutrino signals from dark matter annihilation in the Sun via metastable mediators}",
    eprint = "1102.2958",
    archivePrefix = "arXiv",
    primaryClass = "hep-ph",
    doi = "10.1088/1475-7516/2011/04/003",
    journal = "JCAP",
    volume = "04",
    pages = "003",
    year = "2011"
}

@article{Feng:2015hja,
    author = "Feng, Jonathan L. and Smolinsky, Jordan and Tanedo, Philip",
    title = "{Dark Photons from the Center of the Earth: Smoking-Gun Signals of Dark Matter}",
    eprint = "1509.07525",
    archivePrefix = "arXiv",
    primaryClass = "hep-ph",
    reportNumber = "UCI-TR-2015-07",
    doi = "10.1103/PhysRevD.93.015014",
    journal = "Phys. Rev. D",
    volume = "93",
    number = "1",
    pages = "015014",
    year = "2016",
    note = "[Erratum: Phys.Rev.D 96, 099901 (2017)]"
}

@article{Feng:2016ijc,
    author = "Feng, Jonathan L. and Smolinsky, Jordan and Tanedo, Philip",
    title = "{Detecting dark matter through dark photons from the Sun: Charged particle signatures}",
    eprint = "1602.01465",
    archivePrefix = "arXiv",
    primaryClass = "hep-ph",
    reportNumber = "UCI-TR-2016-02",
    doi = "10.1103/PhysRevD.93.115036",
    journal = "Phys. Rev. D",
    volume = "93",
    number = "11",
    pages = "115036",
    year = "2016",
    note = "[Erratum: Phys.Rev.D 96, 099903 (2017)]"
}

@article{Allahverdi:2016fvl,
    author = "Allahverdi, Rouzbeh and Gao, Yu and Knockel, Bradley and Shalgar, Shashank",
    title = "{Indirect Signals from Solar Dark Matter Annihilation to Long-lived Right-handed Neutrinos}",
    eprint = "1612.03110",
    archivePrefix = "arXiv",
    primaryClass = "hep-ph",
    reportNumber = "MI-TH-1632, WSU-HEP-1608, LA-UR-16-29252",
    doi = "10.1103/PhysRevD.95.075001",
    journal = "Phys. Rev. D",
    volume = "95",
    number = "7",
    pages = "075001",
    year = "2017"
}

@article{Leane:2017vag,
    author = "Leane, Rebecca K. and Ng, Kenny C. Y. and Beacom, John F.",
    title = "{Powerful Solar Signatures of Long-Lived Dark Mediators}",
    eprint = "1703.04629",
    archivePrefix = "arXiv",
    primaryClass = "astro-ph.HE",
    doi = "10.1103/PhysRevD.95.123016",
    journal = "Phys. Rev. D",
    volume = "95",
    number = "12",
    pages = "123016",
    year = "2017"
}

@article{Kouvaris:2012dz,
    author = "Kouvaris, Chris and Tinyakov, Peter",
    title = "{(Not)-constraining heavy asymmetric bosonic dark matter}",
    eprint = "1212.4075",
    archivePrefix = "arXiv",
    primaryClass = "astro-ph.HE",
    doi = "10.1103/PhysRevD.87.123537",
    journal = "Phys. Rev. D",
    volume = "87",
    number = "12",
    pages = "123537",
    year = "2013"
}

@article{MacGibbon:1991tj,
    author = "MacGibbon, Jane H.",
    title = "{Quark and gluon jet emission from primordial black holes. 2. The Lifetime emission}",
    reportNumber = "LHEA-91-001",
    doi = "10.1103/PhysRevD.44.376",
    journal = "Phys. Rev. D",
    volume = "44",
    pages = "376--392",
    year = "1991"
}

@article{Adarsha:2026tpb,
    author = "Adarsha, H. A. and Chakraborty, Chandrachur and Bhattacharyya, Sudip",
    title = "{Transmutation timescales for the dark matter induced collapse of compact stars into black holes}",
    eprint = "2608.10594",
    archivePrefix = "arXiv",
    primaryClass = "astro-ph.HE",
    doi = "10.1103/zb1m-762n",
    journal = "Phys. Rev. D",
    volume = "114",
    number = "6",
    pages = "063007",
    year = "2026"
}

@article{Zantedeschi:2024ram,
    author = "Zantedeschi, Michael and Visinelli, Luca",
    title = "{Ultralight black holes as sources of high-energy particles}",
    eprint = "2410.07037",
    archivePrefix = "arXiv",
    primaryClass = "astro-ph.HE",
    doi = "10.1016/j.dark.2025.102034",
    journal = "Phys. Dark Univ.",
    volume = "49",
    pages = "102034",
    year = "2025"
}

@article{Boccia:2025hpm,
    author = "Boccia, Andrea and Iocco, Fabio",
    title = "{Could the KM3{\textendash}230213A event be caused by an evaporating primordial black hole?}",
    eprint = "2502.19245",
    archivePrefix = "arXiv",
    primaryClass = "astro-ph.HE",
    doi = "10.1103/qxcj-fpwn",
    journal = "Phys. Rev. D",
    volume = "112",
    number = "6",
    pages = "063045",
    year = "2025"
}

@book{Kiritsis:2019npv,
    author = "Kiritsis, Elias",
    title = "{String Theory in a Nutshell: Second Edition}",
    isbn = "978-0-691-15579-1, 978-0-691-18896-6",
    publisher = "Princeton University Press",
    address = "USA",
    month = "4",
    year = "2019"
}

@article{Cicoli:2023opf,
    author = "Cicoli, Michele and Conlon, Joseph P. and Maharana, Anshuman and Parameswaran, Susha and Quevedo, Fernando and Zavala, Ivonne",
    title = "{String cosmology: From the early universe to today}",
    eprint = "2303.04819",
    archivePrefix = "arXiv",
    primaryClass = "hep-th",
    doi = "10.1016/j.physrep.2024.01.002",
    journal = "Phys. Rept.",
    volume = "1059",
    pages = "1--155",
    year = "2024"
}

@book{Becker:2006dvp,
    author = "Becker, K. and Becker, M. and Schwarz, J. H.",
    title = "{String theory and M-theory: A modern introduction}",
    doi = "10.1017/CBO9780511816086",
    isbn = "978-0-511-25486-4, 978-0-521-86069-7, 978-0-511-81608-6",
    publisher = "Cambridge University Press",
    month = "12",
    year = "2006"
}

@book{Binetruy:2006ad,
    author = "Binetruy, P.",
    title = "{Supersymmetry: Theory, experiment and cosmology}",
    year = "2006"
}

@article{Celli:2024cny,
    author = "Celli, Silvia and Peron, Giada",
    title = "{Detection prospects of very and ultra high-energy gamma rays from extended sources with ASTRI, CTA, and LHAASO}",
    eprint = "2403.03731",
    archivePrefix = "arXiv",
    primaryClass = "astro-ph.HE",
    doi = "10.1051/0004-6361/202449837",
    journal = "Astron. Astrophys.",
    volume = "689",
    pages = "A258",
    year = "2024"
}

@article{Moskalenko:2005ng,
    author = "Moskalenko, Igor V. and Porter, Troy A. and Strong, Andrew W.",
    title = "{Attenuation of vhe gamma rays by the milky way interstellar radiation field}",
    eprint = "astro-ph/0511149",
    archivePrefix = "arXiv",
    reportNumber = "SLAC-PUB-11742",
    doi = "10.1086/503524",
    journal = "Astrophys. J. Lett.",
    volume = "640",
    pages = "L155--L158",
    year = "2006"
}

@article{Popescu:2017dgo,
    author = "Popescu, C. C. and Yang, R. and Tuffs, R. J. and Natale, G. and Rushton, M. and Aharonian, F.",
    title = "{A radiation transfer model for the Milky-Way: I. Radiation fields and application to High Energy Astrophysics}",
    eprint = "1705.06652",
    archivePrefix = "arXiv",
    primaryClass = "astro-ph.GA",
    doi = "10.1093/mnras/stx1282",
    journal = "Mon. Not. Roy. Astron. Soc.",
    volume = "470",
    number = "3",
    pages = "2539--2558",
    year = "2017"
}

@article{Sarkar:2026fge,
    author = "Sarkar, Sambo",
    title = "{Constraining dark matter self-interaction from kinetic heating in neutron stars}",
    eprint = "2604.21652",
    archivePrefix = "arXiv",
    primaryClass = "hep-ph",
    doi = "10.1103/fwrs-w77m",
    journal = "Phys. Rev. D",
    volume = "114",
    number = "6",
    pages = "063021",
    year = "2026"
}

@article{Abbasi:2026mak,
    author = "Abbasi, R. and others",
    title = "{IceCube neutrino point-source searches in the direction of the KM3NeT ultra-high-energy event}",
    eprint = "2609.10931",
    archivePrefix = "arXiv",
    primaryClass = "astro-ph.HE",
    month = "9",
    year = "2026"
}

@article{Jaeckel:2010ni,
    author = "Jaeckel, Joerg and Ringwald, Andreas",
    title = "{The Low-Energy Frontier of Particle Physics}",
    eprint = "1002.0329",
    archivePrefix = "arXiv",
    primaryClass = "hep-ph",
    reportNumber = "CPT-10-18, DESY-10-016, IPPP-10-09",
    doi = "10.1146/annurev.nucl.012809.104433",
    journal = "Ann. Rev. Nucl. Part. Sci.",
    volume = "60",
    pages = "405--437",
    year = "2010"
}

@article{Irastorza:2018dyq,
    author = "Irastorza, Igor G. and Redondo, Javier",
    title = "{New experimental approaches in the search for axion-like particles}",
    eprint = "1801.08127",
    archivePrefix = "arXiv",
    primaryClass = "hep-ph",
    doi = "10.1016/j.ppnp.2018.05.003",
    journal = "Prog. Part. Nucl. Phys.",
    volume = "102",
    pages = "89--159",
    year = "2018"
}

@article{DiLuzio:2020wdo,
    author = "Di Luzio, Luca and Giannotti, Maurizio and Nardi, Enrico and Visinelli, Luca",
    title = "{The landscape of QCD axion models}",
    eprint = "2003.01100",
    archivePrefix = "arXiv",
    primaryClass = "hep-ph",
    reportNumber = "DESY 20-036, DESY-20-036",
    doi = "10.1016/j.physrep.2020.06.002",
    journal = "Phys. Rept.",
    volume = "870",
    pages = "1--117",
    year = "2020"
}

@article{Arvanitaki:2009fg,
    author = "Arvanitaki, Asimina and Dimopoulos, Savas and Dubovsky, Sergei and Kaloper, Nemanja and March-Russell, John",
    title = "{String Axiverse}",
    eprint = "0905.4720",
    archivePrefix = "arXiv",
    primaryClass = "hep-th",
    doi = "10.1103/PhysRevD.81.123530",
    journal = "Phys. Rev. D",
    volume = "81",
    pages = "123530",
    year = "2010"
}

@article{Cadamuro:2011fd,
    author = "Cadamuro, Davide and Redondo, Javier",
    title = "{Cosmological bounds on pseudo Nambu-Goldstone bosons}",
    eprint = "1110.2895",
    archivePrefix = "arXiv",
    primaryClass = "hep-ph",
    reportNumber = "MPP-2011-116",
    doi = "10.1088/1475-7516/2012/02/032",
    journal = "JCAP",
    volume = "02",
    pages = "032",
    year = "2012"
}

@article{Depta:2020wmr,
    author = "Depta, Paul Frederik and Hufnagel, Marco and Schmidt-Hoberg, Kai",
    title = "{Robust cosmological constraints on axion-like particles}",
    eprint = "2002.08370",
    archivePrefix = "arXiv",
    primaryClass = "hep-ph",
    reportNumber = "DESY-20-003, DESY 20-003",
    doi = "10.1088/1475-7516/2020/05/009",
    journal = "JCAP",
    volume = "05",
    pages = "009",
    year = "2020"
}

@article{Arbey:2021mbl,
    author = "Arbey, Alexandre and Auffinger, J{\'e}r{\'e}my",
    title = "{Physics Beyond the Standard Model with BlackHawk v2.0}",
    eprint = "2108.02737",
    archivePrefix = "arXiv",
    primaryClass = "gr-qc",
    reportNumber = "CERN-TH-2021-117",
    doi = "10.1140/epjc/s10052-021-09702-8",
    journal = "Eur. Phys. J. C",
    volume = "81",
    pages = "910",
    year = "2021"
}

@article{Atre:2009rg,
    author = "Atre, Anupama and Han, Tao and Pascoli, Silvia and Zhang, Bin",
    title = "{The Search for Heavy Majorana Neutrinos}",
    eprint = "0901.3589",
    archivePrefix = "arXiv",
    primaryClass = "hep-ph",
    reportNumber = "FERMILAB-PUB-08-086-T, NSF-KITP-08-54, MADPH-06-1466, DCPT-07-198, IPPP-07-99",
    doi = "10.1088/1126-6708/2009/05/030",
    journal = "JHEP",
    volume = "05",
    pages = "030",
    year = "2009"
}

@article{Bondarenko:2018ptm,
    author = "Bondarenko, Kyrylo and Boyarsky, Alexey and Gorbunov, Dmitry and Ruchayskiy, Oleg",
    title = "{Phenomenology of GeV-scale Heavy Neutral Leptons}",
    eprint = "1805.08567",
    archivePrefix = "arXiv",
    primaryClass = "hep-ph",
    doi = "10.1007/JHEP11(2018)032",
    journal = "JHEP",
    volume = "11",
    pages = "032",
    year = "2018"
}

@article{Beltran:2021hpq,
    author = "Beltr{\'a}n, Rebeca and Cottin, Giovanna and Helo, Juan Carlos and Hirsch, Martin and Titov, Arsenii and Wang, Zeren Simon",
    title = "{Long-lived heavy neutral leptons at the LHC: four-fermion single-N$_{R}$ operators}",
    eprint = "2110.15096",
    archivePrefix = "arXiv",
    primaryClass = "hep-ph",
    reportNumber = "FTUV-21-1026.4809, IFIC/21-40",
    doi = "10.1007/JHEP01(2022)044",
    journal = "JHEP",
    volume = "01",
    pages = "044",
    year = "2022"
}

@article{Beltran:2024twr,
    author = "Beltr{\'a}n, Rebeca and Bolton, Patrick D. and Deppisch, Frank F. and Hati, Chandan and Hirsch, Martin",
    title = "{Probing heavy neutrino magnetic moments at the LHC using long-lived particle searches}",
    eprint = "2405.08877",
    archivePrefix = "arXiv",
    primaryClass = "hep-ph",
    doi = "10.1007/JHEP07(2024)153",
    journal = "JHEP",
    volume = "07",
    pages = "153",
    year = "2024"
}

@article{Cholis:2026wug,
    author = "Cholis, Ilias and Krommydas, Iason and Carlini, John",
    title = "{Limits on primordial black holes from the extragalactic gamma-ray background; current status and future projections}",
    eprint = "2606.10013",
    archivePrefix = "arXiv",
    primaryClass = "astro-ph.HE",
    month = "6",
    year = "2026"
}

@article{Carlini:2025bki,
    author = "Carlini, John and Cholis, Ilias",
    title = "{GammaPBHPlotter: A public code for calculating the complete Hawking evaporation gamma-ray spectra from primordial black holes}",
    eprint = "2508.19335",
    archivePrefix = "arXiv",
    primaryClass = "astro-ph.HE",
    month = "8",
    year = "2025"
}
\end{document}